# Direct Numerical Simulation of High Prandtl Number Fluid Flow in the Downcomer of an Advanced Reactor

Tri Nguyen* and Elia Merzari

*The Pennsylvania State University, Ken and Mary Alice Lindquist Department of Nuclear Engineering, State College, Pennsylvania*

**Abstract** — *The design of advanced nuclear reactors [Generation IV (Gen IV)] involves an array of challenging fluid-flow issues that affect its safety and performance. Given that Gen IV designs have improved passive safety features, the downcomer plays a crucial role in loss-of-power scenarios. Fluid-flow behavior in the downcomer can involve forced to mixed to natural convection, and characterizing the heat transfer for these changing regimes is a daunting challenge. The creation of a high-resolution heat transfer numerical database can potentially support the development of precise and affordable reduced-resolution heat transfer models. These models can be designed based on a multiscale hierarchy developed as part of the recently U.S. Department of Energy–funded Center of Excellence for Thermal Fluids Applications in Nuclear Energy, which can help address industrial-driven issues associated with the heat transfer behavior of advanced reactors. In this paper, the downcomer is simplified to heated parallel plates, and high Prandtl number fluid (FLiBe) is considered for all simulations. The calculations are performed for a wide range of Richardson numbers from 0 to 400 at two different FLiBe Prandtl numbers (12 and 24), which result in 40 simulated cases in total. Time-averaged and time series statistics, as well as Nusselt number correlations, are investigated to illuminate mixed convection behavior. The calculated database will be instrumental in understanding flow behavior in the downcomer. Ultimately, we aim to evaluate existing heat transfer correlations, and some modifications are proposed.*

**Note** — *Some figures may be in color only in the electronic version.*

## I. INTRODUCTION

Turbulence modeling of heat transfer historically has been very challenging, even for traditional coolants like water. Modern reactor designs often use nontraditional coolants, which have heat transfer characteristics where traditional water-based correlations (e.g., Prandtl unity heat transfer correlations) cannot be applied.

Among nontraditional coolants, molten salts are promising coolants for advanced reactor designs thanks to their special thermophysical and neutronic properties.[1] For example, FLiBe, a mixture of lithium fluoride (LiF) and beryllium fluoride ($BeF_2$), is one of the most well-studied molten salts, as was successfully demonstrated in the Molten Salt Reactor Experiment[2] from 1965 to 1969 at Oak Ridge National Laboratory. However, a comprehensive heat transfer characteristics database for molten salts is still not yet established, as traditional water-based correlations may not be applicable. Currently, there is a significant gap in direct numerical simulation (DNS) data that can be used to advance reactor design. Moreover, the experimental results and high-fidelity numerical simulations, which could be used to better understand the physics, are ineffectively implemented.

Recently, the Center of Excellence for Thermal Fluids Applications in Nuclear Energy[3] (the Center) has been established as a collaboration between Idaho National Laboratory and Argonne National Laboratory.

---

*E-mail: nguyen.tri@psu.edu

A university consortium led by The Pennsylvania State University has been established as a component of the Center to develop and validate reliable and affordable advanced thermal-hydraulic models. The consortium has identified turbulence modeling for heat transfer as an industry-driven challenge problem. In particular, mixed convection effects in the downcomer of advanced reactors have been identified as part of a challenging problem driven by industrial needs.

We aim to generate high-fidelity data for a broad range of nondimensional parameters. The database, driven by requirements identified by our industry partner Kairos Power (KP), will be used to illuminate the flow behaviors in the downcomer, and finally, it will be used to develop reduced resolution models and evaluate existing heat transfer correlations.

In this paper, we will discuss DNS performed to investigate mixed convection heat transfer in the downcomer for a wide range of Richardson numbers at two different FLiBe Prandtl (Pr) numbers (12 and 24) at Re numbers from 5000 to 10 000. The spectral element code NekRS (Ref. 4) is used for all simulation cases thanks to its superior calculation speed. NekRS is a novel refactor of Nek5000 (Ref. 5) oriented toward use in GPU architectures that has demonstrated its capability to perform high-fidelity large eddy simulation (LES) and DNS in both simple and complex geometries.[6–8] The downcomer can be simplified as flow between two heated parallel plates due to the high aspect ratio of most configurations. The working fluid is FLiBe, which is the interest of our industrial partner KP. A wide range of Ri ranging from 0 (forced convection) to 400 has been investigated. Forced convection case data have been compared with the available DNS data of Kasagi,[9] Kasagi et al.,[10] and Kasagi and Lida,[11] and excellent agreement has been achieved.

It is important to note that for high Pr fluid, the momentum diffusivity dominates the thermal diffusivity, and the momentum boundary layer is thicker than the thermal boundary layer. The Batchelor length scale must be used to estimate the resolution requirements; it can characterize the diffusion of heat by molecular processes, which cannot be achieved by using the Kolmogorov length scale. The Batchelor length scale is the ratio of the Kolmogorov length scale to the square root of the Schmidt number:

$$\eta_T = \frac{\eta}{Sc^{\frac{1}{2}}} .$$

For heat transfer, the Pr number can be used instead of the Schmidt number:

$$\eta_T = \frac{\eta}{Pr^{\frac{1}{2}}} = \left(\frac{\alpha^2 \nu}{\epsilon}\right)^{\frac{1}{4}} .$$

The need to resolve the Batchelor length scale everywhere for high Pr fluids like FLiBe means that the spatial resolution can be orders of magnitude higher than for low Pr fluids.[12] This, in turn, results in a drastically increased computational cost. Kawamura et al.[13] could perform DNS of Pr fluid up to 5, at $Re_\tau$ = 180, a significant achievement given the limitation of computational power at that time. In their work, turbulence heat flux (THF) budgets and temperature variance were obtained and visualized for a channel with symmetric heating boundary conditions (BCs).

Statistical correlation coefficients for Pr > 5 were also proposed for future efforts. As computational power improves, more works are performed with in-depth turbulence statistics analyses for low Pr numbers at higher Re numbers, especially for airflow. However, DNS works for high Pr fluid greater than 10 at high Re numbers (Re ≥ 5000) are still very limited.

You et al.[14] performed DNS for heated vertical air flows in fully developed turbulent mixed convection conditions at Re = 2650. Their simulation results show that turbulence statistics profiles for velocity and temperature strongly change when the heat flux increases for both up and downward flows, and the Nu number increases with increasing heat flux. The difference in external and structural effects at different heating conditions in upward heat flow is also revealed.

Recently, Tai et al.[15] conducted a series of DNSs for low-to-unity Pr number fluids at Re = 5000 ($Re_\tau \approx 160$). Detailed turbulence statistics data, such as turbulence kinetic energy (TKE) and THF budgets, as well as time signal and frequency characteristics, were investigated and analyzed. The Nusselt number dataset for low-to-unity Pr number fluid is also provided. Flageul et al.[16] applied conjugate heat transfer (CHT) models for channel flow using wall-resolved LES at high $Re_\tau$ ranging from 150 to 1020 for low-to-unity Pr number fluids. Their CHT LES results were compared with the DNS works of Tiselj and Cizelj,[17] and good agreement in estimation of the temperature variance and scalar dissipation rate have been achieved. They also proposed correlations for regression for the discontinuity of the scalar dissipation rate at higher $Re_\tau$ and different Prandtl numbers, but still, they need to perform more simulations for validation. The authors also noted that the wall-resolved CHT LES models are thoroughly validated for Pr ≪ 1.

Traditionally, Reynolds-Averaged Navier-Stokes (RANS) turbulence models have been applied to overcome the computational cost of DNS. However, the use of RANS turbulence models can impose significant uncertainty due to the restrictive and limiting assumptions of their formulation. Most RANS models were developed based on shear-generated turbulence mechanisms, and they are all suited for mixed convection applications. However, additional terms must be introduced to account for buoyancy-produced turbulence, and these corrections require adequate data for testing and validation.[18,19]

Wang et al.[20] attempted to use Launder-Sharma and Chien RANS models to reproduce velocity, turbulence, and heat transfer characteristics made in the experiments. However, some significant discrepancies between the measurements and predictions are observed in the case when buoyancy opposes the flow. They also noted that in their study, the direct gravitational source term is absent in the transport equations for TKE and dissipation rate, which may be the reason for the disagreement between RANS and experimental results. Furthermore, most of the RANS models of common application rely on the Boussinesq eddy viscosity concept[21] and employ a linear relationship between the anisotropic stresses and the velocity strain rate tensor. Because of the intrinsic limitations in RANS models, advanced turbulence models are often implemented and compared with high-fidelity validation data sets, which are typically provided by DNS data.

Kim et al.[22] performed RANS for fully developed airflow in a vertical pipe at a Reynolds number of 5300 and compared it with DNS data for the same conditions. Their simulation results show that buoyancy is the dominant mechanism that causes laminarization and deterioration of heat transfer. This phenomenon is also observed in this work throughout the budgets and Nusselt number calculation results. The work of Kim et al. also evaluated the performance of the various turbulence models.

Dehoux et al.[23] attempted to apply the elliptic blending (EB) strategy with differential flux models (DFMs) for RANS to improve the predictions of the dynamic and thermal fields from forced to natural convection for turbulent channel flow. The agreement between EB-DFMs with DNS can be drastically improved with substitution of a mixed timescale for the mechanical timescale in the buoyancy production term of the dissipation equation, especially for the natural convection regime. The authors also show that other RANS models can better predict the flow behavior in the natural convection regime as soon as the mix timescale is used in the turbulent dissipation equation, which is also the most important modification in EB-DFMs.

The DNS database in this work will be generated for parallel plates with a focus on differential heating/ cooling applied to case 1 and symmetric cooling applied to case 2. Unlike many previous cases, the flow is allowed to fully develop, which enables one to investigate flow development effects. Figure 1 shows samples of velocity and temperature fields of cases at Pr = 12 and 24. The DNS can capture the pulsating behavior of case 2, as shown in the velocity field of Fig. 1a. The same behavior is also observed on the cold wall side of case 2. However, at Pr = 24, the pulsation intensity is reduced due to the Prandtl effect, as shown in the velocity field of Fig. 1b. The interactions between hot and cold plumes of case 2 are well captured in the temperature field of Fig. 1b.

The computational setups of cases 1 and 2 will be described in detail in Sec. II. In Sec. III, we discuss validation by comparing it with Kasagi DNS data. Sections IV, V, and VI will focus on the results and discussion where the calculated time-averaged statistics, such as the TKE and THF budgets, have been analyzed. Furthermore, snapshots of proper orthogonal decomposition (POD) introduced by Sirovich[24] have been implemented for better understanding and interpreting of the flow behaviors. Time signal and frequency analyses have also been considered, and differences in power spectral density (PSD), wavelet energy content, and frequency between cases 1 and 2 have been revealed. The Nusselt number calculations have been estimated for each case, and trends have been compared against existing experimental data. This analysis shows good agreement between numerically estimated Nusselt numbers and available correlations as well as experimental data. We are also proposing novel correlations for high Pr flows as they are transitioning from mixed convection to natural convection. Section VII summarizes the paper and provides some conclusions.

## II. NUMERICAL METHODOLOGY

### II.A. Numerical Method

The highly scalable high-order spectral element code NekRS is chosen for both LES and DNS thanks to its capability to perform high-fidelity LES and DNS for not merely canonical but also complex geometry problems[25–27] at superior calculation speed compared to its predecessor Nek5000. NekRS takes advantage of all features in Nek5000 and utilizes them effectively in GPU architectures, thus drastically improving the calculation speed compared to Nek5000. In NekRS, the GPU kernels are written in the Open Concurrent Compute Abstraction (OCCA) library, which provides portability among different parallel languages such as OpenCL, CUDA, and HIP. Users can use OCCA to implement the parallel kernel code in C++ language and OKL to convert Nek5000 cases to NekRS ones.

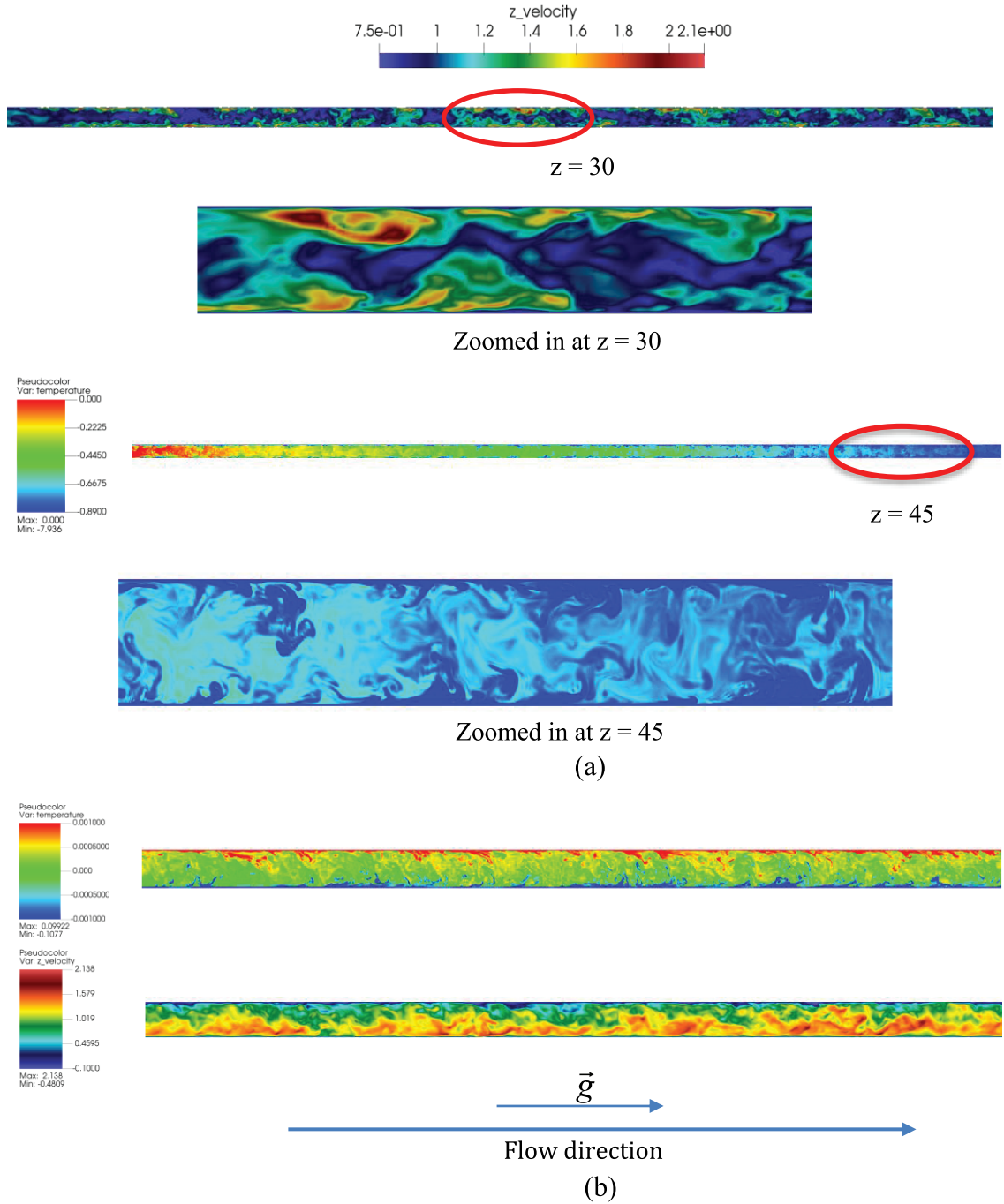

Fig. 1. An example of velocity and temperature fields of symmetric cooling parallel plates with mixed convection. (a) Case 2, Pr = 12 and (b) case 1, Pr = 24.

Both Nek5000 and NekRS can address cases with variable thermophysical properties using low-Mach number approximation.[4,28] The main difference between low-Mach compressible Navier-Stokes (NS) equations and incompressible NS equations is the non-zero divergence term in the continuity equation, which accounts for the change in density. However, in this work, we considered only incompressible NS equations with constant properties Newtonian fluid (FLiBe) subject to buoyancy, modeled through the Boussinesq approximation:

*Mass continuity equation*:

$$\nabla \cdot u = 0 ; \qquad (1)$$

*Momentum conservation equation*:

$$\rho\left(\frac{\partial u}{\partial t} + u \cdot \nabla u\right) = -\nabla P + \mu \nabla \cdot \nabla u + \rho g(1 - \beta(T - T_0)) ; \quad (2)$$

*Energy conservation equation expressed in terms of temperature*:

$$\rho c_p \left(\frac{\partial T}{\partial t} + u \cdot \nabla T\right) = \nabla \cdot (k \nabla T) , \quad (3)$$

where

- $u, g$ = vector velocity and gravity, respectively
- $\rho$ = density of fluid
- $\mu$ = dynamic viscosity
- $\beta$ = heat expansion coefficient
- $k$ = conductivity
- $c_p$ = heat capacity.

The NS equations are solved using the spectral element method (SEM). The SEM, introduced by Patera,[29] is a subclass of the Galerkin methods. It is a high-order method that provides good geometric versatility comparable to finite element methods but also with minimum numerical dispersion/dissipation and the rapid convergence properties typical of spectral methods.[5] Nek5000 is highly scalable[30,31] and has been used in many works to gain new insight into the physics of turbulence in complex flows.

In SEM, the domain is discretized into $E$ hexahedral elements and represents the solution as a tensor-product of $N$'th-order Lagrange polynomials based on Gauss-Lobatto-Legendre (GLL) nodal points, which results in $E(N + 1)^3$ degrees of freedom (DOF) per scalar field. GLL points are chosen for efficient quadrature for both velocity and pressure spaces. The pressure can be solved at the same polynomial order of the velocity $N$ ($P_N - P_N$ formulation) or at lower order $N - 2$ ($P_N - P_{N-2}$ formulation). Two time-stepping schemes, both up to the third order, are available: backward differentiation formula and operator-integration factor scheme.[32] The latter has the advantage of less severe stability limitation allowing for Courant–Friedrichs–Lewy condition (CFL) > 1, but it results in a higher cost per time step.

We simulate 40 cases in this work. The nondimensionalnon-dimensional form of NS.equations (1), (2), (3) is applied:

$$\nabla \cdot u^* = 0 , \quad (4)$$

$$\left(\frac{\partial u^*}{\partial t^*} + u^* \cdot \nabla u^*\right) = -\nabla P^* + \nabla \cdot \frac{1}{\text{Re}} \nabla u^* - \text{Ri} \times T^*, \quad (5)$$

and

$$\rho^*\left(\frac{\partial T^*}{\partial t^*} + u^* \cdot \nabla T^*\right) = \nabla \cdot \frac{1}{\text{Pe}} \nabla T^*, \quad (6)$$

where

- $u^*$ = nondimensional velocity = $u/U$, where $U$ = inlet bulk velocity
- $t^*$ = time = $t/(D/U)$, where $D = 4\delta$, where $D$ = hydraulic diameter and $\delta$ = half-channel width
- $\rho^*$ = density = $\rho/\rho_0 = 1$
- $P^*$ = pressure = $P/(\rho_0 U^2)$, where $\rho_0$ = inlet density
- $T^*$ = temperature = $(T - T_0)/\Delta T$, where $T_0$ = inlet temperature and $\Delta T$ = temperature difference between inlet and outlet (for case 2 in Fig. 2b) or between the two walls (for case 1 in Fig. 2b)
- Re = Reynolds number
- Ri = Richardson number
- Pe = Peclet number.

Note that Re = $(DU\rho_0)/\mu$; Fr = $U/\sqrt{gD}$; Pe = PrRe = $(\rho_0 UDc_{p0})/k$; and Ri = $\beta\Delta T/\text{Fr}^2$, where Fr is the Froude number.

In order to characterize the mixed convection phenomena, specifically, the relative strength between forced and natural convection, the Richardson number Ri quantitatively defined the force term in Eq. (5); Ri is defined by the ratio of the Grashof number, Eq. (7), to the square of the Re number, as shown in Eq. (8):

$$\text{Gr} = \frac{\rho^2 g \beta \Delta T D^3}{\mu^2} \quad (7)$$

and

$$\text{Ri} = \frac{\text{Gr}}{\text{Re}^2} = \frac{\beta \Delta T}{\text{Fr}^2} . \quad (8)$$

## II.B. Numerical Setup

The computational model for the parallel plate case is shown in Fig. 2a. The fluid domain is divided into three sections: upper adiabatic, heat flux, and lower adiabatic. The ratio among them, as well as the spanwise $L_x$ and wall-normal $L_y$ lengths, is defined according to the interest of our industrial partner KP:

$$L_x = \pi L_y , \qquad (9)$$

$$L_{Z0} = L_{Z3} = 10 L_y , \qquad (10)$$

$$L_{Z1} = L_{Z0} + L'_{Z0} = 10L_y + 2L_y = 12L_y, \quad (11)$$

and

$$L_{Z2} = 100L_y. \quad (12)$$

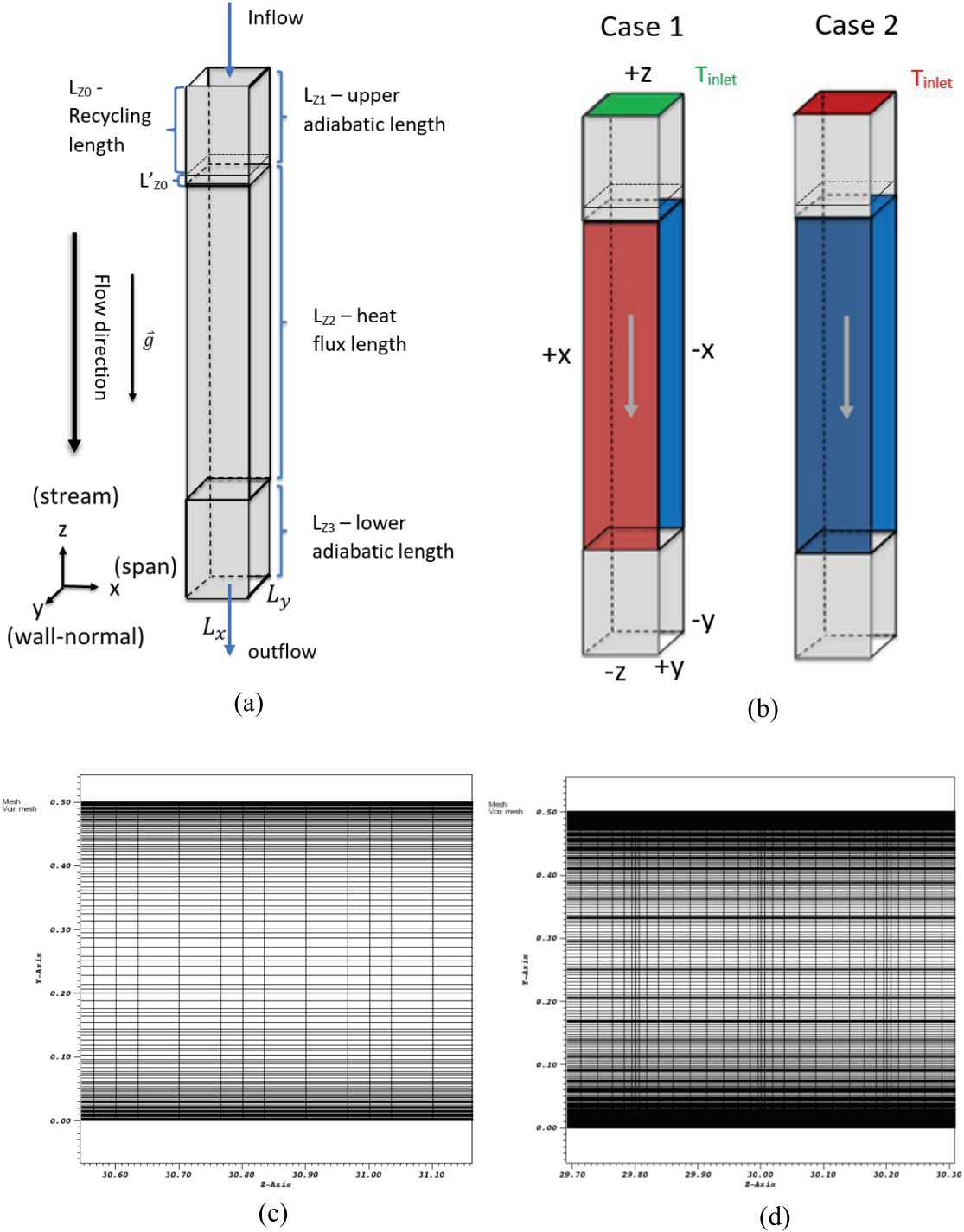

Fig. 2. (a) The simplified downcomer is presented as parallel plate computational box. (b) Boundary condition map for investigated cases. (c) and (d) The grid employed in the simulation from half of the channel's cross section for Re = 5000 at (c) the 7th order and (d) the 10th order.

According to Eqs. (9) through (12), the streamwise length of the channel is $122L_y$. To achieve a fully developed turbulent velocity profile at desired Re before entering the heat flux section, a recycling method is applied with the length of $L_{Z0}$. Note that for the upper adiabatic length, $L_{Z1} = L_{Z0} + L'_{Z0}$, where $L'_{Z0}$ is the distancing between the recycling length and the heat flux length, which equals 1D. The gravity is not applied in the upper adiabatic length to ensure that the fully developed turbulence flow generated by recycling enters the heated length properly. The length of $L_x$ was determined after an extensive sensitivity study. The TKE budgets for $L_x = \pi L_y$, $2\pi L_y$, and $4\pi L_y$ were unchanged for the mixed convection cases, leading to the conclusion that $L_x = \pi L_y$ is a sufficient length.

The Gmsh open-source three-dimensional finite element mesh generator[33] has been used to generate the mesh. The resulting mesh has 250 000 hexahedra. All simulation cases use the same mesh but different polynomial orders. For instance, case 1 at Re = 10 000 with Pr = 12 required 11th polynomial order ($P_N$), which results in 677 million DOF. The discretization has satisfied standards for the DNS of channel flow, which includes $y+ < 1$ near the wall, ten grid points below $y+ = 10$, $\Delta x+ < 4$, and $\Delta z+ < 8$, at Pr = 1. We note that the resolution is significantly higher. We also investigated mesh convergence by comparing simulation results for the 7th, 9th, and 10th polynomial orders (as shown in Figs. 2c and 2d), and we observed no meaningful differences in generated turbulence data. The time step in the simulations is set to a maximum of $10^{-3}$ s (CFL < 0.5), and the gravitational direction was chosen to be on the z-axis, as shown in Fig. 2a.

We investigated two classes of cases (case 1 and case 2), as shown in Fig. 2b, representing different cooling conditions at the wall boundary. The working fluid is FLiBe at 695°C and 560°C, which corresponds to Pr = 12 and 24 in nondimensional terms. The Neuman BC imposing constant heat flux is applied for two opposite vertical walls in $+y$ and $-y$ directions. The spanwise directions $+x$ and $-x$ are periodic. The Dirichlet BCs imposing constant temperature are used at the inlet. The heat flux applied to each heated plate is also nondimensionalized; it is defined by Eq. (13):

$$f^* = f \cdot \left(\frac{1}{\rho_0 U c_p \Delta T}\right), \quad (13)$$

where $f^*$ and $f$ are nondimensional heat flux and dimensional heat flux, respectively.

For case 1, the temperature difference $\Delta T$ between the walls is defined by Eq. (14):

$$\Delta T = \frac{f \times D}{2k}. \quad (14)$$

Hence, the nondimensional heat flux will become

$$|f^*| = |f| \cdot \left(\frac{1}{\rho_0 U c_p \Delta T}\right) = \frac{2k}{\rho_0 U c_p D} = \frac{2}{\text{RePr}} \\ = \frac{2}{\text{Pe}}, \quad (15)$$

and Ri from Eq. (8) will become

$$\text{Ri} = \frac{g\beta |f| D^2}{2kU^2}. \quad (16)$$

For case 2, the temperature difference $\Delta T$ between the inlet and the outlet, the nondimensional heat flux, and the Ri number are defined by Eqs. (17), (18), and (19), respectively:

$$\Delta T = \frac{|Q|}{\dot{m} c_p} = \frac{f \times 2L_{Z2}}{\rho_0 U c_p L_y}, \quad (17)$$

$$|f^*| = |f| \cdot \left(\frac{1}{\rho_0 U c_p \Delta T}\right) = \frac{L_y}{2L_{Z2}}, \quad (18)$$

and

$$\text{Ri} = \frac{g\beta |f| D}{\rho_0 U^3 c_p}, \quad (19)$$

where $Q$ is the heat energy and $\dot{m}$ is the mass flow rate.

The initial conditions are the same for all cases:

$$T^* = 0 \quad (20)$$

and

TABLE I

the Velocity and Temperature BCs Applied for Case 1 and Case 2

| Boundary Condition Geometry | Velocity Boundary Conditions | | Temperature Boundary Conditions | |
|---|---|---|---|---|
| | Case 1 | Case 2 | Case 1 | Case 2 |
| $+x$ and $-x$ | Periodic | Periodic | Periodic | Periodic |
| $+y$, for $L_{Z1}$ and $L_{Z3}$ | No-slip | No-slip | Isolated | Isolated |
| $+y$, for $L_{Z2}$ | No-slip | No-slip | Heat flux | Heat flux |
| $-y$, for $L_{Z1}$ and $L_{Z3}$ | No-slip | No-slip | Isolated | Isolated |
| $-y$, for $L_{Z2}$ | No-slip | No-slip | Heat flux | Heat flux |
| $+z$ | Inflow | Inflow | $T^* = 0$ | $T^* = 0$ |
| $-z$ | Pressure outflow | Pressure outflow | Isolated | Isolated |

$$u_x^* = u_y^* = u_z^* = 0 \ . \quad (21)$$

The BCs for case 1 and case 2 are summarized in Table I.

The summary of simulation cases and the corresponding mesh resolution are reported in Table II. Each Re number has four corresponding Ri numbers, and two Pr numbers, which lead to eight simulation cases applied for case 1 and case 2. Cases at Re = 10 000 and Pr = 24 are not being performed due to the limitation in computational resources. The conditions listed have been driven by input from our industry partner KP, and they are relevant to KP Fluoride salt-cooled High-temperature Reactor (KP-FHR) accident and operational conditions.

The advantage of using NekRS on GPUs over Nek5000 has been evaluated by comparing the computing time for 1000 time steps for case 1 at Re = 1000 and Ri = 400 on the Summit supercomputer using five compute nodes. Nek5000 completed 1000 time steps in around 1800 s (≈1.8 s/time step) on 120 IBM POWER9 CPU cores, whereas NekRS completed 1000 time steps in just 20s on 30 NVIDIA Volta V100 GPUs (≈0.02 s/time step), demonstrating superior calculation speed toward Nek5000.

## III. VALIDATION

Because of the lack of relevant references for the simulation conditions of mixed convection cases (Ri ≠ 0), as referenced in Table II, a separate simulation campaign with the forced convection case (Ri = 0) has been performed to compare with Kasagi et al.[9–11] DNS data. The Re and Pr values have been selected to match the Kasagi et al. cases. We note that in the work of Kasagi et al., the hydraulic diameter is defined by $D = 2\delta$, where $\delta$ is the half-channel width. In our work, $D = 4\delta$, which makes the Re number increase by twice

TABLE II

the Simulation Cases with Their Corresponding Reynolds, Grashof, and Richardson Numbers and Mesh Resolution

| Reynolds | Cases | | | | Mesh Resolution | | | |
|---|---|---|---|---|---|---|---|---|
| | Case 1 | | Case 2 | | Pr = 12 | | Pr = 24 | |
| | Richardson | Grashof | Richardson | Grashof | $P_N$ | DOF | $P_N$ | DOF |
| 5000 | 0 | 0 | 0 | 0 | 7 | 172 million | 9 | 365 million |
| 5000 | 8 | $2 \times 10^8$ | 0.04 | $10^6$ | | | | |
| 5000 | 80 | $2 \times 10^9$ | 0.4 | $10^7$ | | | | |
| 5000 | 400 | $10^{10}$ | 2 | $5 \times 10^7$ | | | | |
| 7500 | 0 | 0 | 0 | 0 | 9 | 365 million | 12 | 866 million |
| 7500 | 3.56 | $2 \times 10^8$ | 0.0178 | $10^6$ | | | | |
| 7500 | 35.56 | $2 \times 10^9$ | 0.178 | $10^7$ | | | | |
| 7500 | 177.78 | $10^{10}$ | 0.889 | $5 \times 10^7$ | | | | |
| 10 000 | 0 | 0 | 0 | 0 | 11 | 677 million | Not performed | Not performed |
| 10 000 | 2 | $2 \times 10^8$ | 0.01 | $10^6$ | | | | |
| 10 000 | 20 | $2 \times 10^9$ | 0.1 | $10^7$ | | | | |
| 10 000 | 100 | $10^{10}$ | 0.5 | $5 \times 10^7$ | | | | |

in comparison with Kasagi et al., turbulence statistic parameters including average streamwise velocity $\langle u \rangle^+$, root-mean-square (rms) streamwise velocity $u^+_{rms}$, average local temperature difference $\langle \theta \rangle^+$, and rms local temperature difference $\theta^+_{rms}$, as well as TKE budgets in the streamwise direction, are compared with previous DNS data, and excellent agreement has been achieved as shown on Figs. 4 and 5. It is noted that the turbulence statistics parameters are plotted against the $y^+$ axis. $Y^+$ is the nondimensional wall distance. It can be defined by the following:

$$y^+ = \frac{y u_\tau}{v} ,$$

where $y$ is the dimensional wall distance, $u_\tau$ is the friction velocity, and $v$ is the kinematic viscosity.

The use of $y^+$ in Figs. 4 and 5 is consistent with the nondimensional approach. In computational fluid dynamics, $y^+$ is often used to describe the mesh quality for particular flow patterns. It is important to reach the desired $y^+$ in turbulence modeling for the proper size of the cells near domain walls. In our DNS high Pr fluid calculations, $y^+$ needs to satisfy the Batchelor length scale, making $y^+$ much smaller than 1, while the Kolmogorov length scale typically requires $y^+ \leq 1$. TKE budget is a concept used to understand and predict the behavior of turbulent flows. It helps to determine the sources and sinks of TKE, which can have a significant impact on the overall behavior of the flow. In particular, TKE is the mean kinetic energy associated with eddies in turbulence flow. TKE can be transferred down to the turbulence energy cascade and is dissipated by viscous forces at the Kolmogorov scale. The process of production, transport, and dissipation of TKE can be expressed by the TKE budget equation in Fig. 3. The definition of $y^+$ and TKE budgets can also be found in Ref. 34.

The THF budgets are also compared with previous DNS data, and excellent agreement has also been achieved. The comparison results are delivered in detail in Refs. 35 and 36. Note that the agreement with Kasagi DNS data could only be archived with specific run-time settings, which are then implemented for all cases in Table II.

## IV. TIME-AVERAGED STATISTICS OF HEATED PARALLEL PLATES AT DIFFERENT RICHARDSON, REYNOLDS, AND PRANDTL NUMBERS

### IV.A. Turbulence Statistics Analyses

Before collecting data, all cases have been run for at least three flows through time to reach statistical convergence, as recommended by run-time settings acquired from our validation campaign. The effect of buoyancy for cases with Ri numbers ranging from 0.04 to 400 at different Re (5000, 7500, and 10 000) and Pr (12 and 24) numbers is investigated and analyzed. A series of comparison campaigns has been conducted for all cases, including normalized averaged and rms of velocity and temperature, TKE, and THF. All the turbulence data presented in this section have been collected on a z-plane cross section at $z = L_{Z1} + 0.5 L_{Z2} = 62 L_y$, where the flow is considered fully developed. Note that for case 1, the turbulence statistics, including average, rms, TKE, and THF budgets of velocity and temperature, were collected for both the hot wall and the cold wall due to the asymmetric temperature BCs, which leads to opposite signs of buoyancy forces at the hot wall and cold wall sides. For case 2, because of the symmetric temperature BCs, the statistics are the same for both sides of the wall. We also compared the turbulence statistics results for case 1 at Re = 5000, Pr = 12 at different spanwise lengths ($L_x = \pi/2$, $\pi$, and $2\pi$), and no changes in turbulence statistics have been observed.

First, the effect of buoyancy for cases at the same Reynolds and Prandtl numbers but different Richardson

$$\frac{D\bar{k}}{Dt} = -\frac{1}{\rho_0} \frac{\overline{\partial u_i' p'}}{\partial x_i} - \frac{1}{2} \underbrace{\frac{\overline{\partial u_j' u_j' u_i'}}{\partial x_i}}_{T_K} + \underbrace{v \frac{\partial^2 \bar{k}}{\partial x_j^2}}_{D_K} - \underbrace{\overline{u_i' u_j'} \frac{\partial \bar{u}_i}{\partial x_j}}_{P_K} - \underbrace{v \frac{\overline{\partial u_i'}}{\partial x_j} \frac{\partial u_i'}{\partial x_j}}_{\ni_K} - \frac{g}{\rho_0} \overline{p' u_i'} \delta_{i3}$$

| Parameter | Physical meaning |
|---|---|
| TKE (Turbulence Kinetic Energy) $\bar{k}$ | Mean kinetic energy associated with turbulence eddies in the flow |
| TKE dissipation $\ni_K$ | Turbulence dissipation rate |
| TKE production $P_k$ | TKE growth rate in the flow |
| TKE turbulence diffusion $T_k$ | The transport of TKE due to turbulence itself |
| TKE viscous diffusion $D_k$ | The transport of TKE due to molecular viscosity |

Fig. 3. TKE budget equation and TKE budget physical meaning.

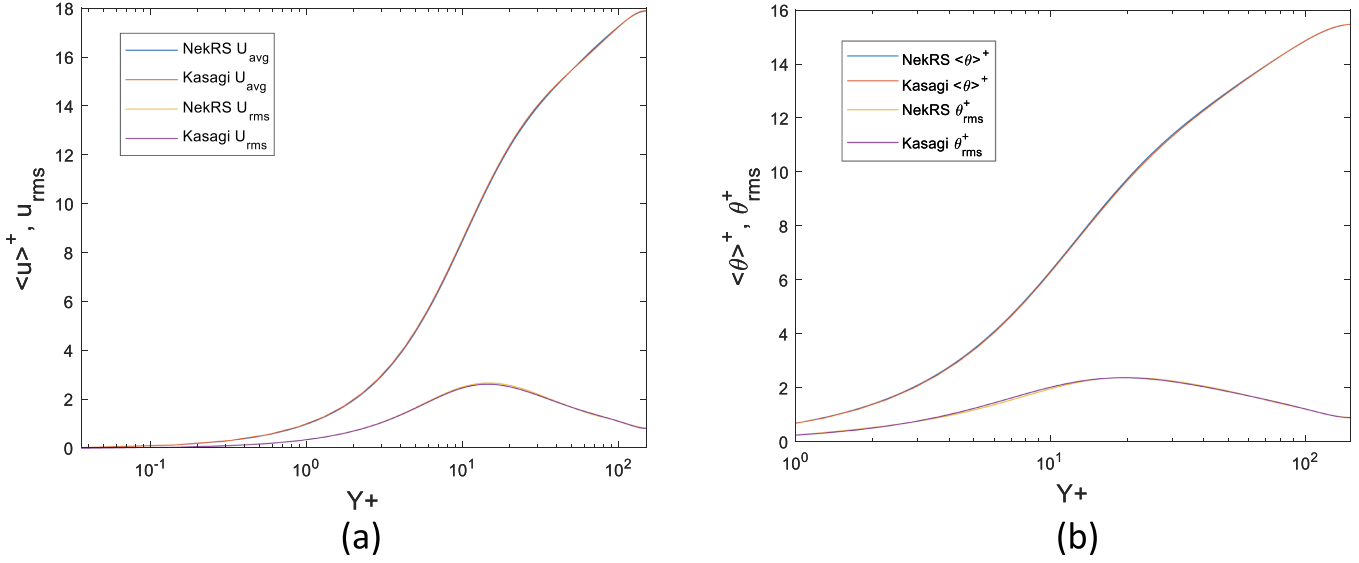

Fig. 4. (a) Average and rms streamwise velocity and (b) average and rms local temperature difference of NekRS and Kasagi et al.

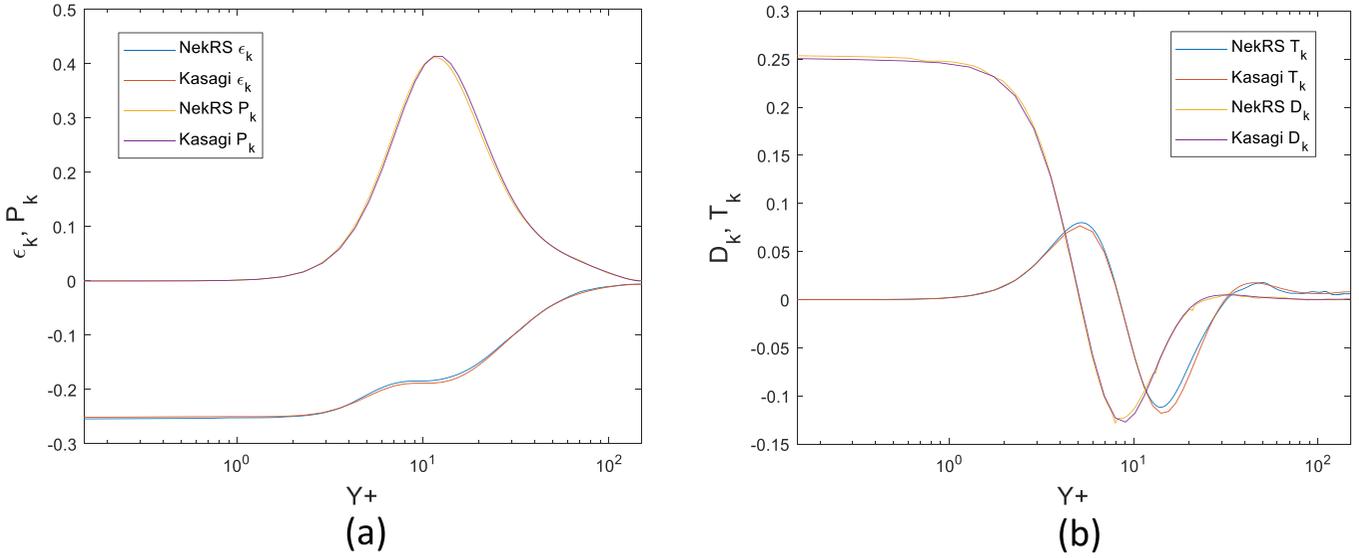

Fig. 5. TKE budgets: (a) dissipation term $\epsilon_k$ and production term $P_k$ and (b) viscous diffusion term $D_k$ and turbulence diffusion term $T_k$ of NekRS and Kasagi et al.

numbers from 0.04 to 400 is investigated and analyzed for Re = 5000 and Pr = 12. For other Re and Pr datasets in Table II, the behavior is similar and not presented here in the interest of brevity. A comparison campaign of turbulence statistics has been conducted between forced convection cases (Ri = 0) and mixed convection cases (Ri ≠ 0) for both case 1 and case 2 (Figs. 6 and 7). In general, the average velocity and temperature profiles, as well as the TKE budgets, dramatically change as the Ri number increases and the flow transitions from forced to natural convection. The behavior of turbulence profiles in case 2 is similar to that of the cold wall side in case 1. For case 1, the rms of velocity ($u^+_{rms}$) and temperature ($\theta^+_{rms}$) of the cold wall side is lower than that of the hot wall side, as shown in 6b, 6e, 6h, 6j, 6k, and 6l.

In fact, on the cold wall side, the buoyancy force accelerates the flow whereas on the hot wall side, it pushes against the flow, leading to increased dissipation, high TKE levels, and chaotic flow structure. We note that it is typically referred to as downward flow in mixed convection literature. The effect of inertial forces is particularly evident in the production term $P_k$. Case 2 shows a shift from a single peak (forced convection) to a double peak (mixed convection), as shown in Fig. 7h, which indicates that massive $P_k$ happens not only close to the wall but also near the center of the channel for mixed convection. But, with a sufficiently high Ri number, both walls in case 2 and the cold wall in case 1 have regions with negative production $P_k$. This is typical of conditions with strong flow acceleration due to the buoyancy force (Figs. 7b and 7h). The bigger Ri is, the stronger is the negative $P_k$ and the closer it is to the wall. On the other hand, the production of the hot wall for case 1 is significantly higher than the cold wall, which can be explained by the opposite direction of the buoyancy force in this region (Fig. 7e). This also can explain the differences between the hot wall and the cold wall of case 1 for other TKE budgets terms; in particular, the amplitude of dissipation $\epsilon_k$ and the viscous diffusion $D_k$ are about four times larger on the hot wall side in comparison with the cold wall side, as shown in Figs. 7a, 7d, 7g, 7c, 7f, and 7i).

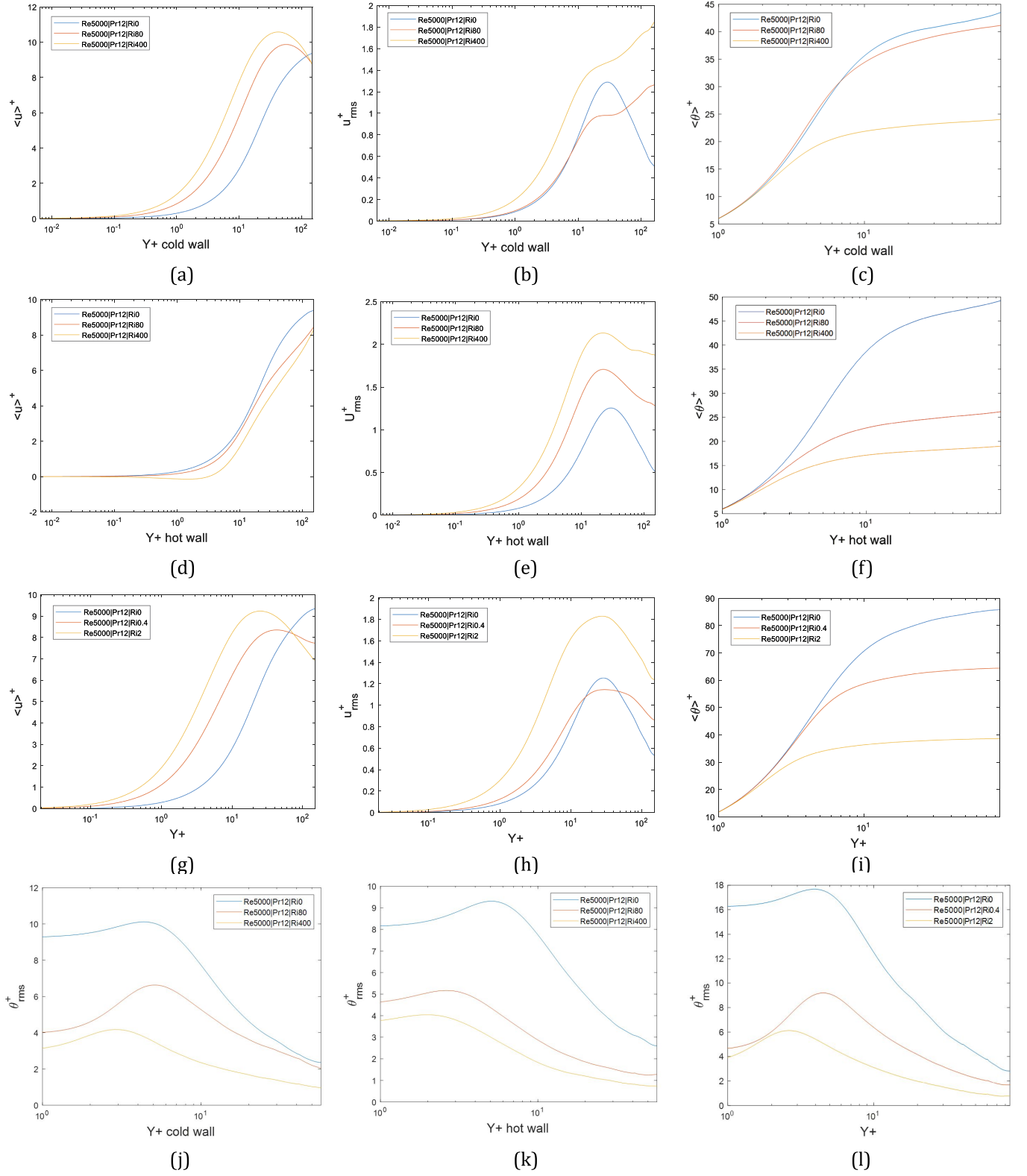

Fig. 6. (a) and (d) Average streamwise velocity $\langle u \rangle^+$ for case 1 (Ri = 0, 80, and 400). (g) $\langle u \rangle^+$ for case 2 (Ri = 0, 0.4, and 2.0). (b) and (e) rms streamwise velocity $u^+_{rms}$ for case 1. (h) $u^+_{rms}$ for case 2. (c) and (f) Average local temperature difference $\langle \theta \rangle^+$ for case 1. (i) $\langle \theta \rangle^+$ for case 2. (j) and (k) rms local temperature difference $\theta^+_{rms}$ for case 1. (l) $\theta^+_{rms}$ for case 2.

The statistics behavior of cases at Re = 7500 and 10 000 are similar to the case at Re = 5000 as the Ri number increases. The THF budgets are also considered to evaluate the case behavior at different Ri numbers. It can be seen in Fig. 8 that the THF budgets are strongly affected as the Ri number changes. We observe a negative THF production region, as shown in Fig. 8b, which is caused by flow acceleration due to the buoyancy force.

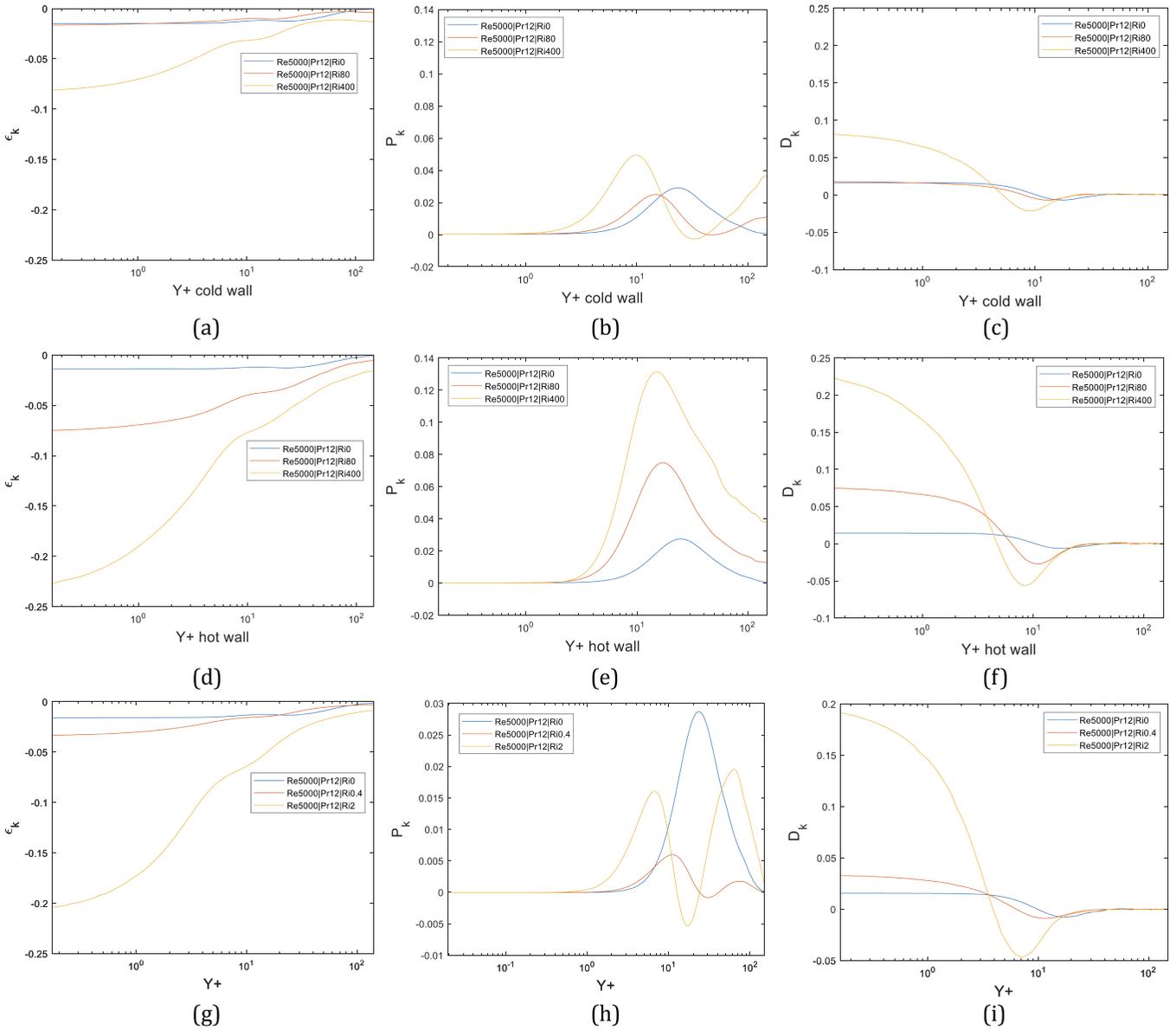

Fig. 7. (a) and (d) TKE dissipation $\epsilon_k$ for case 1. (g) $\epsilon_k$ for case 2. (b) and (e) TKE production $P_k$ for case1. (h) $P_k$ for case 2. (c) and (f) Viscous diffusion $D_k$ for case 1. (i) $D_k$ for case 2.

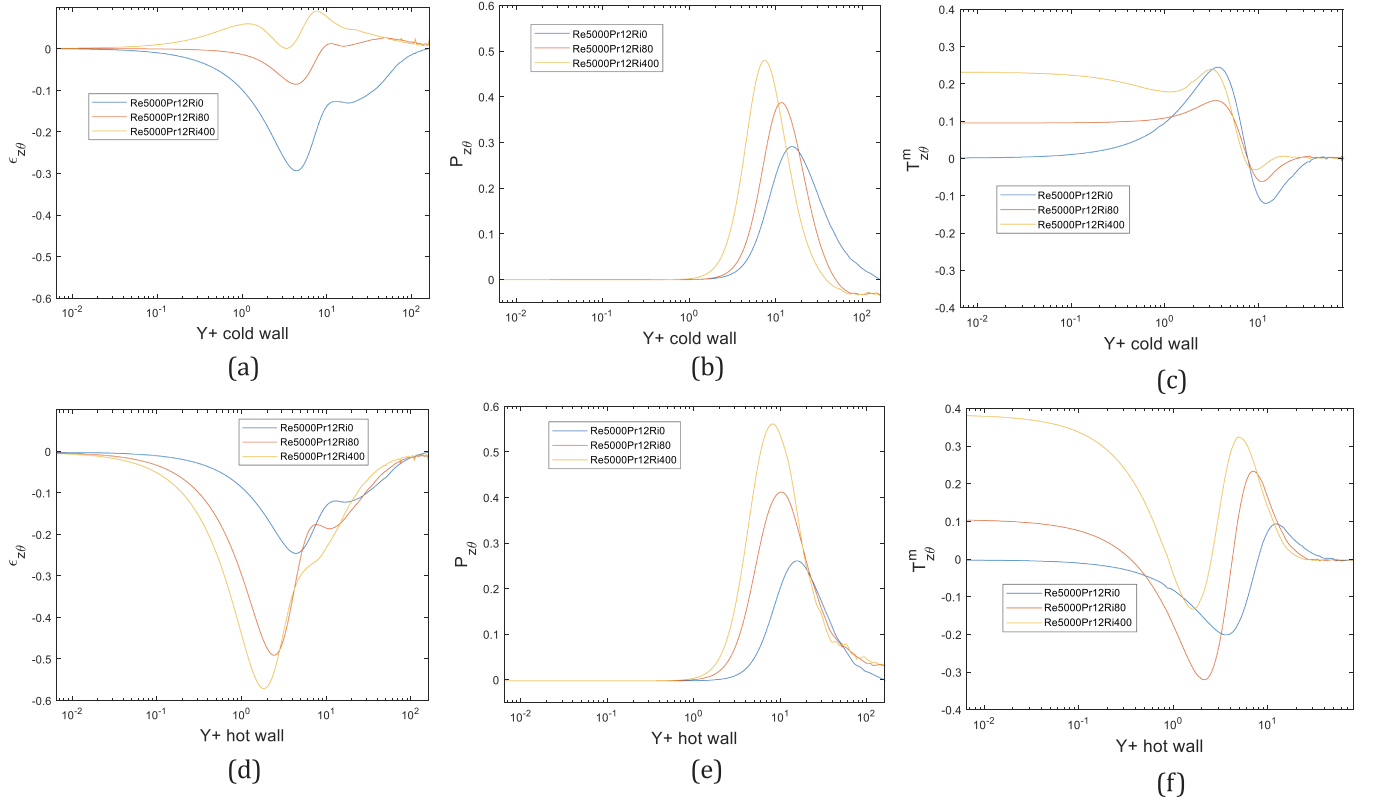

Fig. 8. THF budgets for case 1 at Re = 5000 and Pr = 12. (a) and (d) Streamwise THF dissipation $\epsilon_{z\theta}$ at (a) cooled wall and (d) heated wall. (b) and (e) Streamwise THF production $P_{z\theta}$ at (b) cooled wall and (e) heated wall. (c) and (f) Streamwise THF viscous diffusion $T_{z\theta}^m$ at (c) cooled wall and (f) heated wall.

However, the effects are less pronounced than for the TKE budgets.

The effect of buoyancy for cases at different Prandtl numbers is also investigated. Figure 9 shows the turbulence statistics behavior of several cases at Re = 5000 but different Pr and Richardson numbers. At the same Re number, the larger the Pr number is, the weaker is the the Ri number effect. Overall, the amplitude of turbulence statistics decreases as the Pr number increases. In some conditions, for instance, case 1 at Ri = 80, the flow shifts from a mixed convection case at Pr = 12 to a nearly forced convection case at Pr = 24, and the negative production region vanishes, as shown in Figs. 9b and 9e. For case 2, the Pr effect strongly reduces the TKE production double peak amplitude, thus reducing the intensity of the mixed convection effect. The bigger Ri is, the stronger this effect is, as shown in Figs. 9c and 9f.

Finally, cases at different Reynolds numbers are compared with each other. Figure 10 shows the TKE budget behavior of case 2 at different Re numbers but the same Pr number. It is important to note that cases 1 and 2 in Table II use the same set of Gr numbers applied for each group of Re numbers, which leads to a different set of Ri numbers. For forced convection cases, the higher the Re number is, the stronger the budgets are. That can be explained by enhancement of turbulence intensity at a high Re number. However, while mixed convection cases depend strongly on Ri numbers, Re numbers have little impact on flow behavior. At the same Ri number, the budgets and the average and rms of velocity behave similarly, as shown in Fig. 10.

### IV.B. Budget Peaking Investigation

In this section, statistics at different streamwise locations of the channel have been collected to evaluate changes across the streamwise direction. The data have been gathered at $z = 10, 20, 30, 40,$ and $50$, and sample results are shown in Figs. 11a and 11b. Figures 12a and 12b show the average velocity and temperature for a case where the buoyancy is strong enough to cause a significant deflection and redistribution of the momentum in the cross section, which is also reflected by the TKE dominant region near the hot wall side in Fig. 12c. When this is observed, we also observe that there are budgets peaking at roughly $z = 20$. This occurs in several cases, as shown in Fig. 11b. To better understand the physics behind this phenomenon, POD analyses[7,24] have been implemented for a series of cases with and without budget peaking. The POD is chosen because it is a very effective mathematical method to identify the dominant flow structure, which in turn can reveal the difference in dynamic behavior between cases with and without peaking.

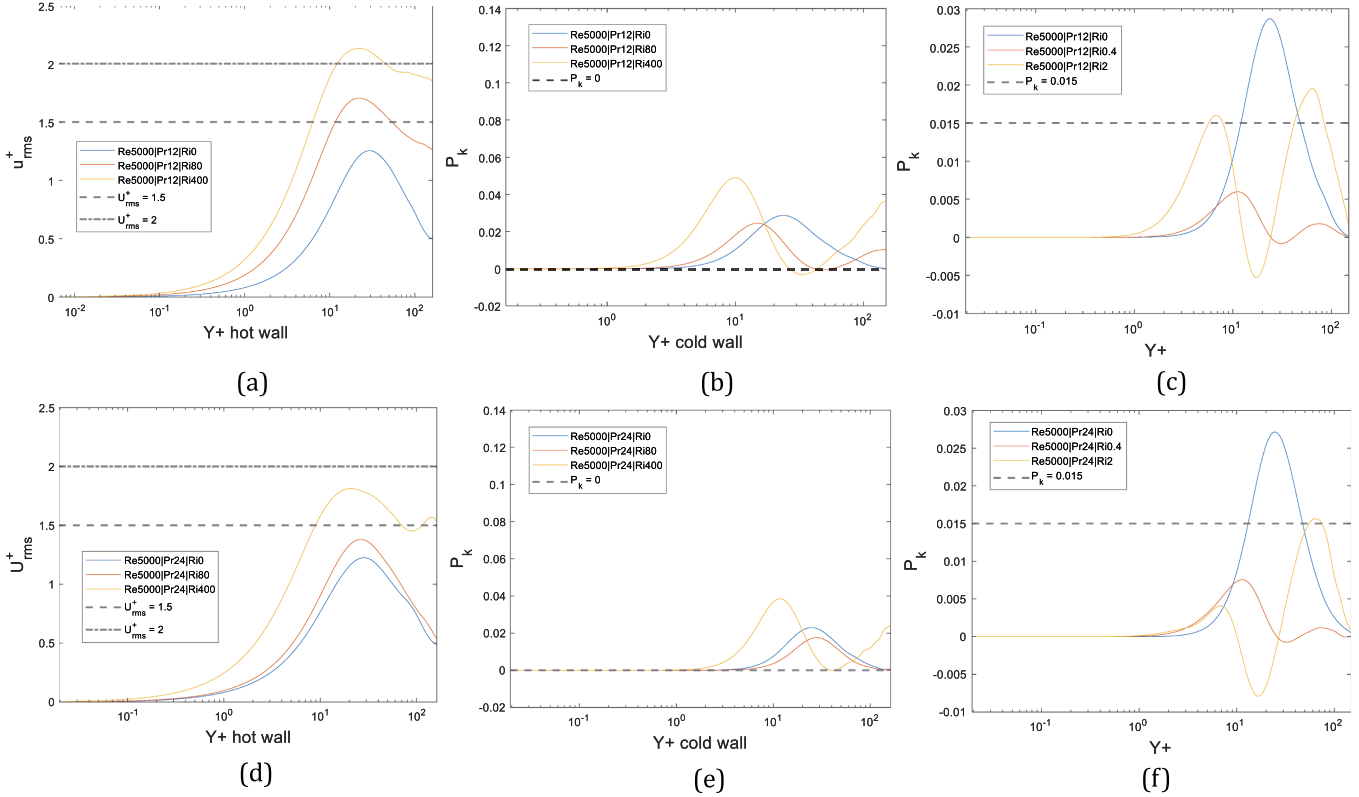

Fig. 9. Turbulence statistic comparison between cases at Pr = 12 and 24 with Re = 5000. (a) and (d) rms of streamwise velocity $u^+_{rms}$ for case 1 at hot wall. (b) and (e) TKE production $P_k$ for case 1 at cold wall. (c) and (f) TKE production for case 2.

To perform POD, 1000 instantaneous snapshots, sufficiently separated in time to avoid excessive correlation, are generated for each simulation case as POD input data. The first three POD modes corresponding to sample cases with and without budgets peaking at $z = 20$ are shown in Fig. 11c. It can be clearly seen that the POD modes of cases with budget peak present significant differences from the non-peaking counterparts. In the region of interest, i.e., $z$ from 13 to 21, the first and second modes of a case at Re = 5000, Pr = 12, and Ri = 400 show high energy content developing from the hot wall. This effect becomes more marked as $z$ increases from 13 to 21. The size of the energy structure continues to grow until it reaches the cold wall, and then it decreases in strength, as shown in Fig. 11c. Such a flow structure cannot be seen in cases at Re = 5000, Pr = 12, and Ri = 80, where the budget peaking is absent. We also observe the same flow behavior for other cases with budget peak, but we do not present it here in the interest of brevity.

## V. TIME SERIES AND FREQUENCY ANALYSIS

### V.A. Time Series Observation

For the purpose of time series assessment, the time signals of streamwise velocity and temperature at Re = 5000 for case 1 (Ri = 0 and Ri = 400) and case 2 (Ri = 0 and Ri = 2.0) are considered. The time series data are collected from points near the wall and points in the middle of the channel at $z = 31$. Figure 13 shows that the oscillation frequency and amplitude of $v_z$ increase drastically from forced (Figs. 13a, 13b, and 13c) to mixed convection cases (Figs. 13d through 13i). We also observe the presence of a low-frequency content as the Ri number increases (case 1 at Ri = 400 and case 2 at Ri = 2.0). We note that the center point of case 1 shows a stronger oscillation amplitude than that of case 2. The large oscillation amplitude of temperature combined with the higher low-frequency content for the points near the wall of the point near the hot wall of case 1 signals the presence of large-scale vortices. This is consistent with what is observed visually near the walls, as shown in Fig. 1. These large vortices are reflected by the temperature PSD peak, which will be shown in the next section. Note that the largest temperature vortices are absent for the forced convection cases and for case 2 in Table II, which is consistent with the absence of PSD peaks for case 2 in the next section.

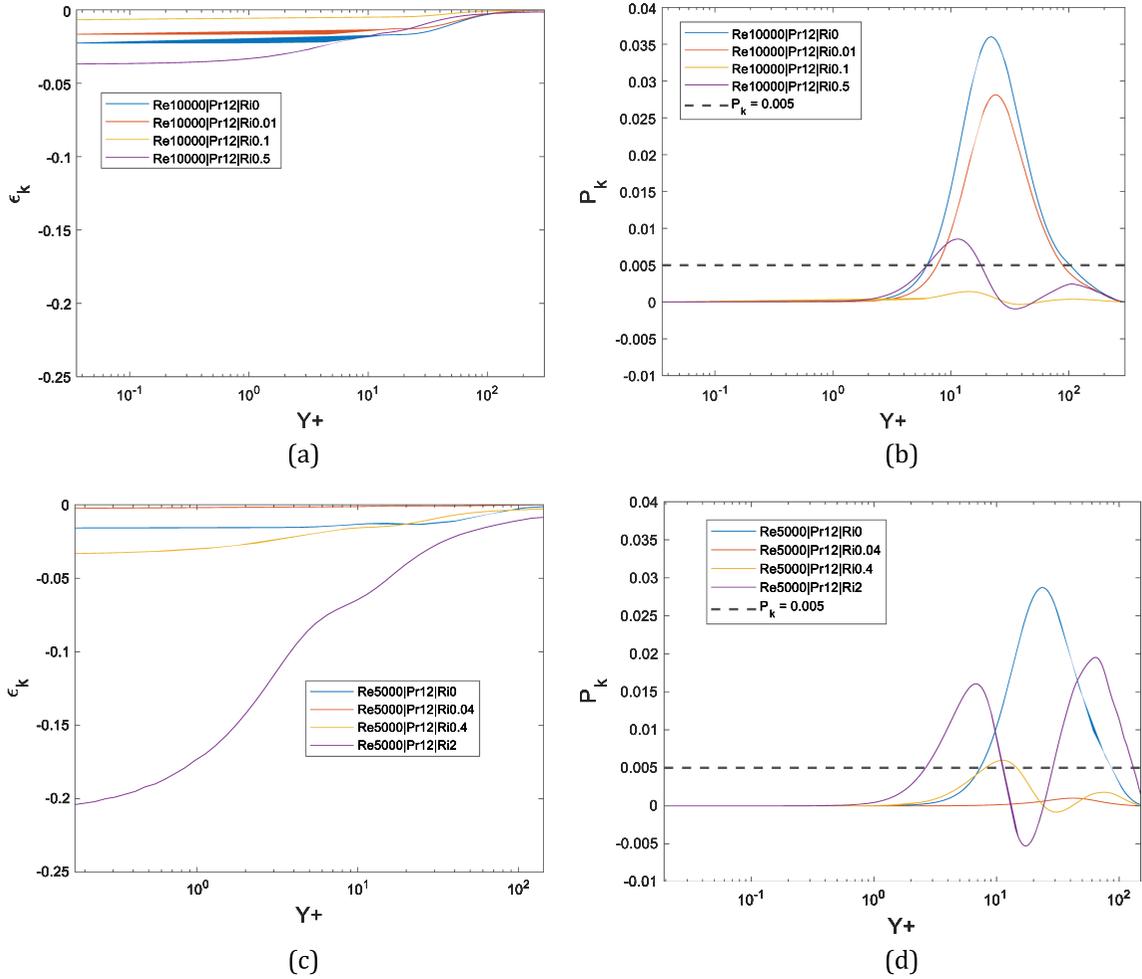

Fig. 10. (a) through (d) TKE budget comparison for case 2 at Pr = 12 with Re = 5000 and 10 000. (a) and (c) TKE dissipation $\epsilon_k$. (b) and (d) TKE production $P_k$.

## V.B. PSD Analyses

In this section, time signals of streamwise velocity and temperature for cases 1 and 2 at different Re and Ri numbers are used to perform PSD analyses. The data are collected from points near the wall and points in the middle of the channel at $z = 31$ for a 1.5 flow through time. Figure 14 shows the PSD of velocity and temperature for Re = 5000 and 10 000 with Pr = 12. In general, high PSD values are observed at low frequencies; the energy content gradually decreases as the frequency increases. For cases at Re = 5000, it can be recognized in Fig. 14c that there are noticeable shifts from low to high frequency as the Ri values increase, which corresponds to the evolution from forced to natural convection. The shift at high frequency occurs between 20 and 70 Hz. There are marked differences in PSD between points in the interior and points near the wall, likely due to the relative difference in the strength of the buoyancy force. Figures 14g, 14h, and 14i point to a peculiar peaking of the temperature PSD, for case 1, Ri = 400 at around 70 Hz on the cold wall side. Based on our observation of temperature and velocity patterns (Sec. IV.B), this peaking is likely to be caused by turbulent structure formation in the early part of the channel (Fig. 11c). In fact, we observe that this PSD peaking occurs only in cases where there is also a budget peaking in the early section of the domain. We also note that the PSD

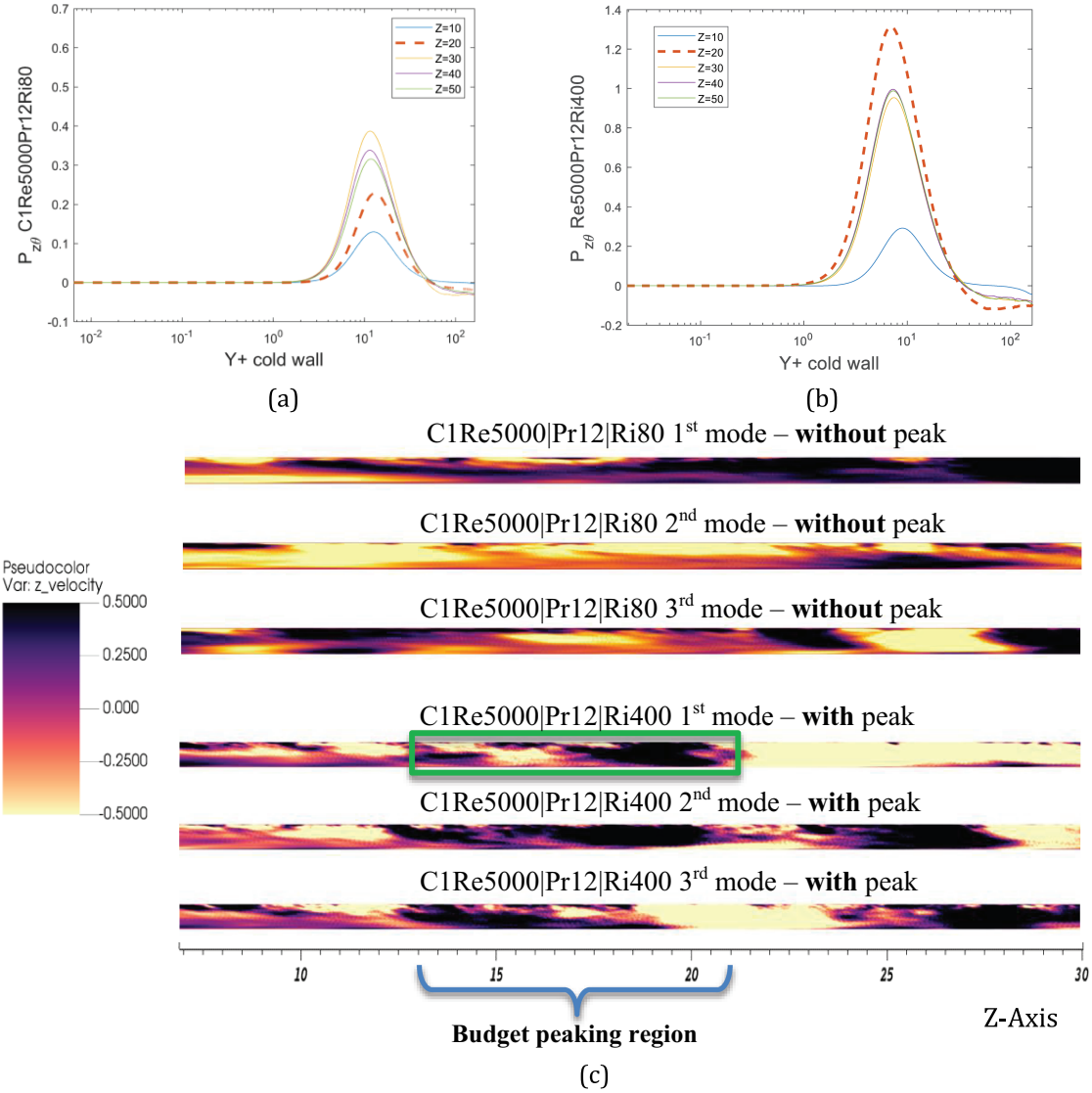

Fig. 11. (a) and (b) THF production of case 1 at Re = 5000, Pr = 12, and Ri = 80 and 400 at multiple streamwise locations on cold wall side. (c) Streamwise velocity POD of case 1 at Re = 5000, Pr = 12, and Ri = 80 and 400.

peaking is not present in case 2. This will be discussed more in Sec. V.C (wavelets).

### V.C. Wavelet Analyses

Wavelet transform is a powerful tool for analyzing time signals. The wavelet transform of the signal is practically a cross correlation of the signal with the mother wavelet at different scales, which can reveal sharp changes in spectral behavior in time and frequency. In general, the wavelet transform is a combination of the Fourier transform and windowed Fourier transform.[37] In this section, the continuous Morlet wavelet[38] is chosen to perform wavelet analyses for the cases in Table I. The time series data are transformed into wavelet functions by Eq. (22):

$$f(t) = \iint \gamma(s,\tau)\psi_{s,\tau}(t) d\tau ds , \qquad (22)$$

where $f(t)$, $\gamma(s,\tau)$, and $\psi_{s,\tau}(t)$ are time series data, coefficients of wavelets, and wavelet function with scale $s$ and time $\tau$, respectively.

For wavelet analyses, the time signals of the cases in Table II have been collected for 122 convective units (1 flow through time). The collection locations are the same as in Sec. V.A. The bright color in the wavelet spectrogram in Figures 15, 16, and 17 shows a high magnitude of the wavelet energy content whereas the opposite applies to the dark color. The wide spectrogram band along the time axis corresponds to a large coherent structure.[39] For the interest of brevity, only prominent wavelet spectrograms are shown in this section.

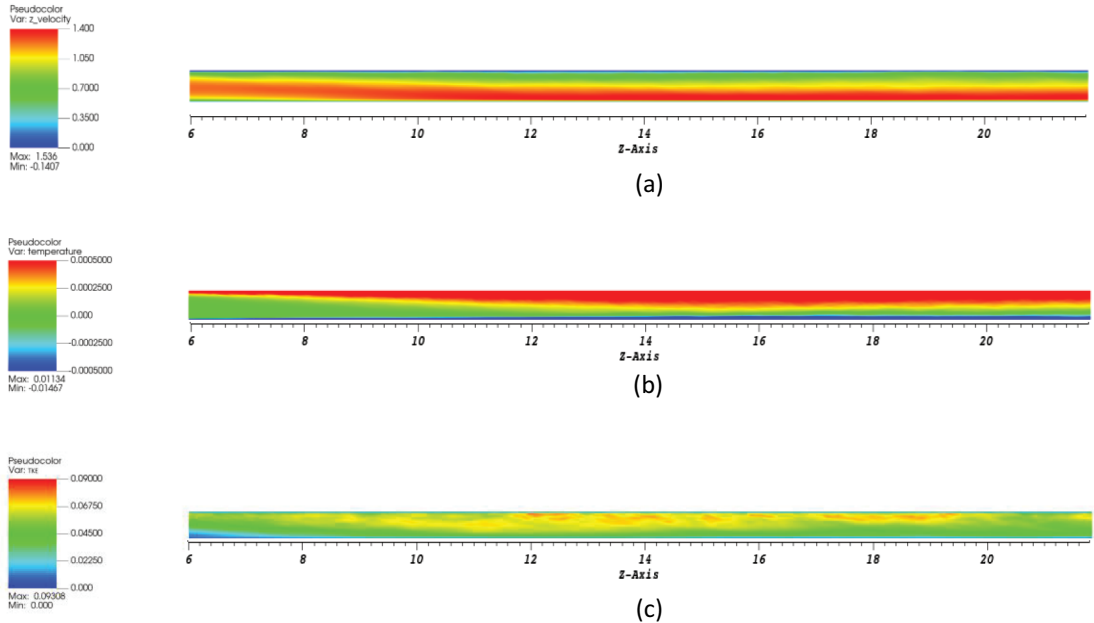

Fig. 12. (a) Average streamwise velocity field and (b) average temperature field of case at Re = 5000, Pr = 12, and Ri = 400.

The buoyancy effect can be clearly seen in Fig. 15. The wavelet energy content increases sharply on the hot wall side from the forced convection case (Ri = 0) to the natural convection–dominated case (Ri = 400), where the buoyancy force strongly goes against the flow on the hot wall side. A large coherent structure also appears in the spectrogram of the case at Ri 400, which is also shown in Fig. 1b. On the cold wall side, the wavelet energy content also increases strongly compared to the forced convection case but less intensely compared to the hot wall side, as shown in Fig. 16. For all cases in Table II, wavelets in the middle of the channel points have lower energy content compared to the near-wall points, which is also reflected in Fig. 1. In general, cases at Pr = 24 have lower wavelet energy content and smaller coherent structures compared to cases at Pr = 12 due to the Prandtl number effect already discussed. For all forced and mixed convection cases, most of the highest energy content is concentrated in the frequency range from 0 to 200 Hz, which corresponds to PSD analysis results. Moreover, the wavelet spectrogram can also reveal the low-frequency behavior between the forced and the mixed convection cases, which cannot be clearly seen in PSD plots.

The wavelet spectrogram for the temperatures of case 1 at Re = 5000, Pr = 12, and Ri = 400 is investigated to better understand the PSD behavior in Sec. V.B. Figure 17 shows that there is a small wavelet structure but has very significant energy content compared to other structures in the spectrogram. The zoomed spectrogram reveals the frequency range of the small structure energy burst from 55 to 70 Hz, which matches the frequency of the PSD peak in Sec. V.B. This type of energy burst is absent in case 2 and other mixed convection cases where PSD peaking is also absent. As proposed in Sec. V.B, the energy burst could be a result of intermittent formation of the turbulent structure in the early part of the channel; the structure grows until reaching the cold wall side where the PSD peaks. The structure then travels downstream, where the wavelet energy burst is observed. This is consistent with the fact that the cases with the PSD peak and wavelet energy burst also have the budgets peaking at the early part of the channel (around z = 20).

## VI. NUSSELT NUMBER DATASET FOR FLIBE

The Nusselt (Nu) number is an important dimensionless number that defines the effectiveness of convection heat transfer. For relatively new nontraditional coolants such as molten salt like FLiBe, there is a lack of reliable heat transfer correlations for a broad range of Grashof and Reynolds numbers. In this work, the Nu number has been calculated for each case in Table II and presented as a function of the Buoyancy number (Bo) (Ref. 40):

$$\text{Bo} = 8 \times 10^4 \frac{\text{Gr}}{\text{Re}^{3.425}\text{Pr}^{0.8}} \quad . \tag{23}$$

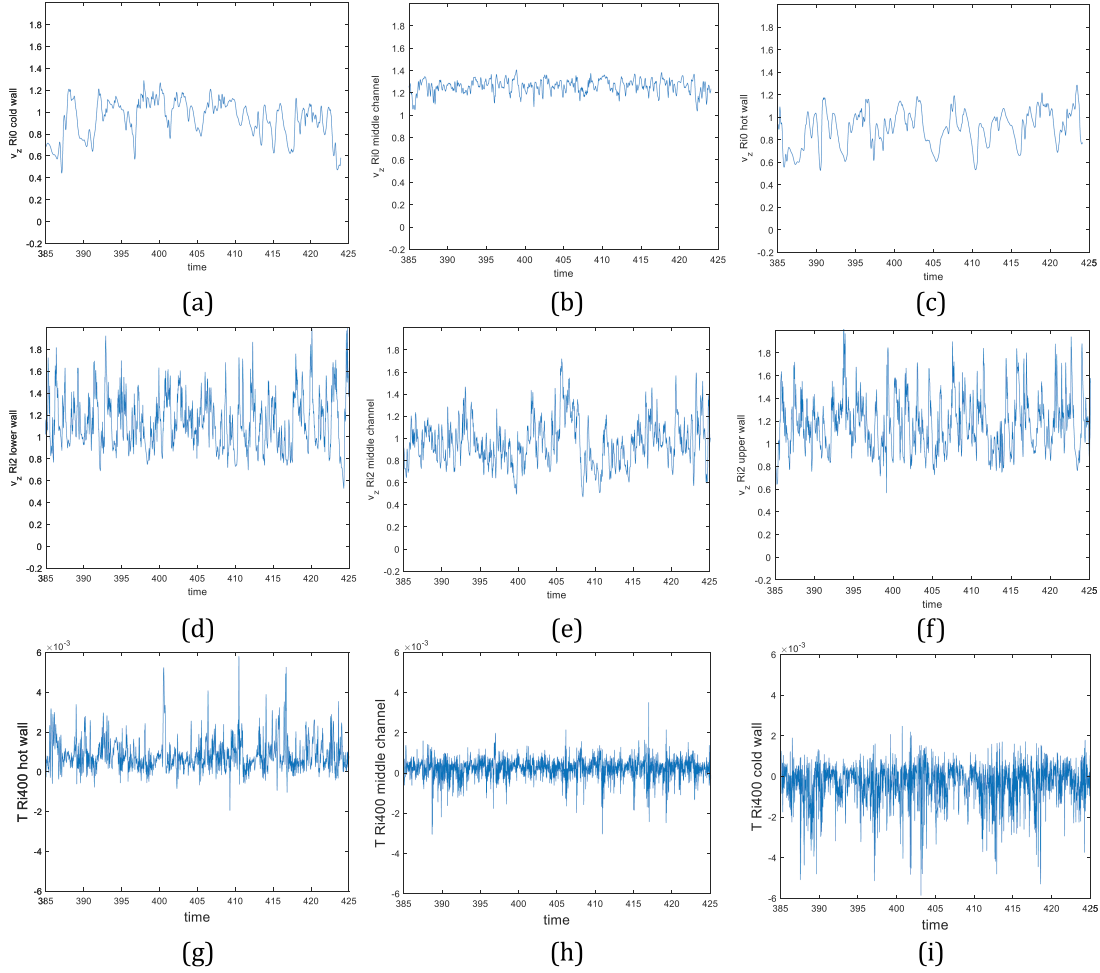

Fig. 13. (a), (b), and (c) Time series of streamwise velocity $v_z(t)$ for forced convection cases. (d), (e), and (f) $v_z(t)$ for case 2 at Ri = 2. (g), (h), and (i) Time series of temperature $T(t)$ for case 1 at Ri = 400.

Overall, the DNS dataset follows the trends of the experimental and KP data.[41] On the hot wall side, the Nu number increases as the Ri number grows, which can be explained by the opposite direction of the increasing buoyancy force enhancing the turbulence. On the cold wall side, the Nu number initially decreases as the Bo number starts to increase from forced convection (Bo = 0) to Bo = 0.3, forming the heat transfer impairment region. Then, the Nu number increases steadily and tends to converge with its counterpart on the hot wall side when Bo is big enough (natural convection). This behavior can be interpreted by the results from Sec. IV.A: Starting from Bo = 0 to 0.3, the buoyancy force starts accelerating the flow, the turbulence flow is laminarized (the TKE production becomes zero or even negative), and the Nu number in turn declines. Following the growth of Bo, the laminarization effect is maximum at around Bo = 0.3, corresponding to the minimum Nu, and the turbulence is slowly reestablished as Bo increases. After that, the Nu number keeps increasing, the natural convection is established for both hot and cold walls, and there is an increasingly smaller difference in the heat transfer coefficient between the two walls.

An effort has been made to find the Nu number correlation describing the trend of the current DNS Nu dataset. A reference correlation from Ref. 40 has been considered for further development:

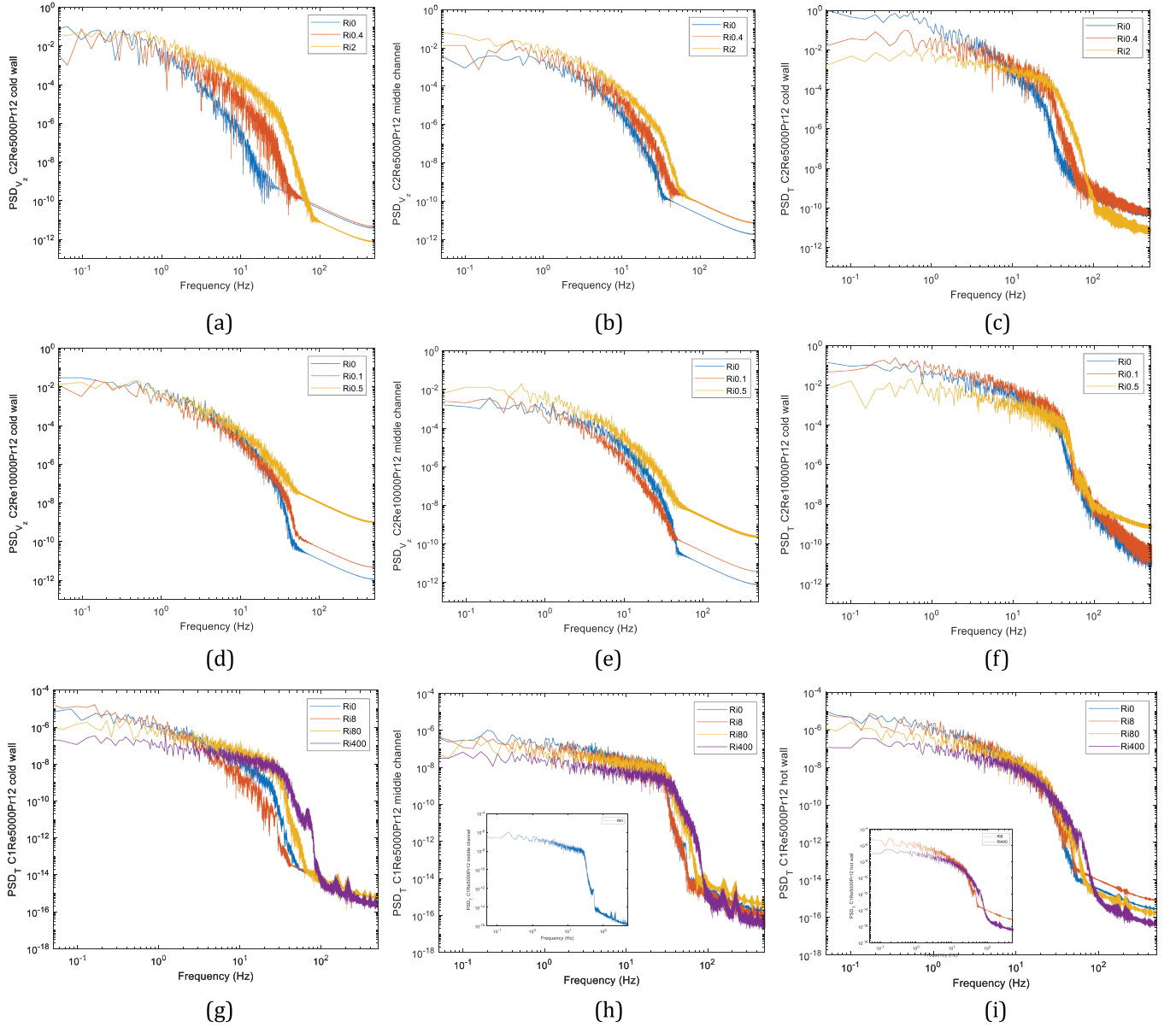

Fig. 14. (a) through (f) PSD of case 2 at Re = 5000 and Re = 10 000 with the same Pr number (Pr = 12). (a) and (d) PSDs of streamwise velocity for a point near wall. (b) and (e) PSDs of streamwise velocity for a point in the middle of the channel. (c) and (f) PSDs of temperature for a point near wall. (g), (h), and (i) PSDs of temperature for case 1.

$$\frac{\text{Nu}}{\text{Nu}_0} = \left(1 \pm \frac{\text{Bo}}{\left(\frac{\text{Nu}}{\text{Nu}_0}\right)^2}\right)^{0.46}, \quad (24)$$

where $\text{Nu}_0$ corresponds to the Nusselt number of the forced convection cases in Table II.

The implementation of Eq. (24) is shown in Fig. 18a. It can be observed that Eq. (24) could not describe the trend of the presented data perfectly. On the cold wall side, it can characterize the dataset behavior in a very limited manner.

A generalized Nu correlation has been offered for a better representation of the DNS, experimental, and KP dataset:

$$\frac{\text{Nu}}{\text{Nu}_0} = \left(1 \pm \frac{\text{Bo}}{\left(a\frac{\text{Nu}}{\text{Nu}_0} + b\right)^2}\right)^c. \quad (25)$$

Applying MATLAB Curve Fitting Toolbox,[42] the new Nu correlation for the current Nu dataset is found as follows:

*For the hot wall side,*

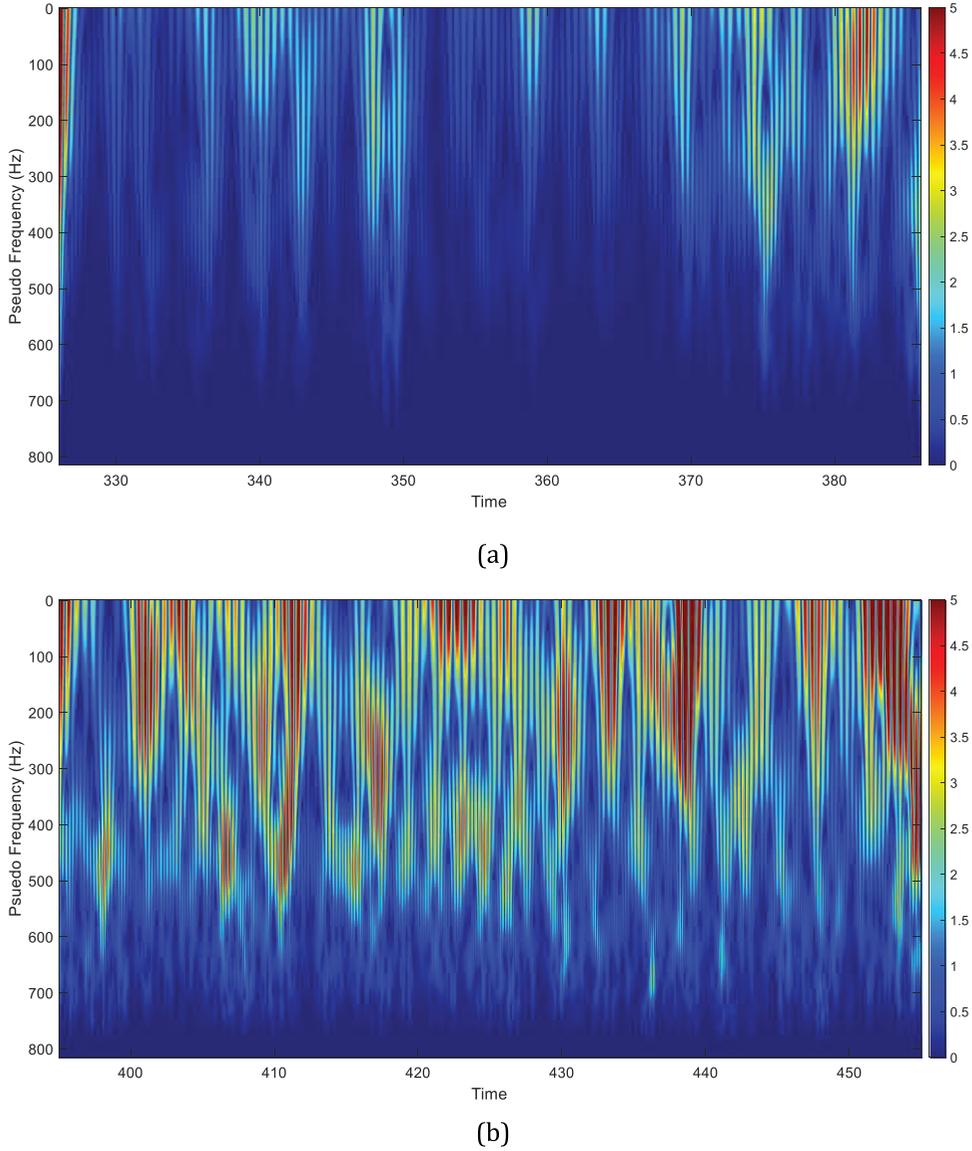

Fig. 15. Wavelet spectrogram for the velocity of case 1 at Re = 5000, Pr = 12, and Ri = 0 (a) and Re = 5000, Pr = 12, and Ri = 400 (b) on the hot wall side.

$$\frac{\text{Nu}}{\text{Nu}_0} = \left(1 + \frac{\text{Bo}}{\left(\frac{\text{Nu}}{\text{Nu}_0}\right)^2}\right)^{0.56}; \quad (26)$$

*For the cold wall side,*

$$\frac{\text{Nu}}{\text{Nu}_0} = \left(1 - \frac{\text{Bo}}{\left(0.56\frac{\text{Nu}}{\text{Nu}_0} + 0.35\right)^2}\right)^{0.56} \quad (27)$$

and

$$\frac{\text{Nu}}{\text{Nu}_0} = \left(1 - \frac{\text{Bo}}{\left(1.78\frac{\text{Nu}}{\text{Nu}_0} - 0.5\right)^2}\right)^{2.03}, \quad (28)$$

where Eq. (27) is applied for Bo values ranging from 0 to 0.29 and Eq. (28) is for Bo > 0.29. The correlations are also presented in Fig. 18a for readability. Figure 18b shows that most of the data points lie on the diagonal within 10% of the confidence interval, which means that Eqs. (26), (27), and (28) can adequately describe the trends of the Nu dataset. The higher the Bo number is, the more accurate the correlations are. There is still Nu data scattering in the heat transfer impairment region that needs to be characterized by functions of other parameters rather than the Bo number, and the solutions of Eqs. (26), (27), and (28) are transcendental, which remains a limitation of the approach that we are trying to overcome with our ongoing work.

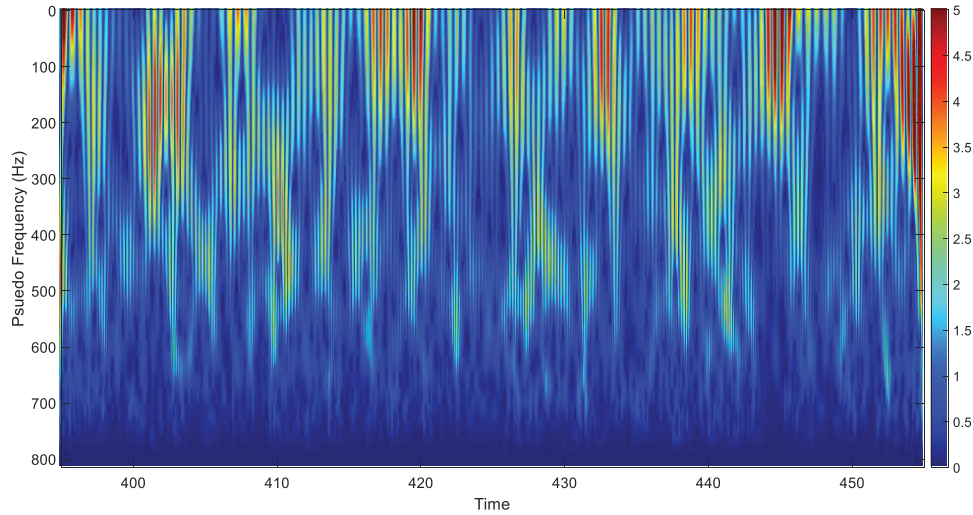

Fig. 16. Wavelet spectrogram for the velocity of case 1 at Re = 5000, Pr = 12, and Ri = 400 on the cold wall side.

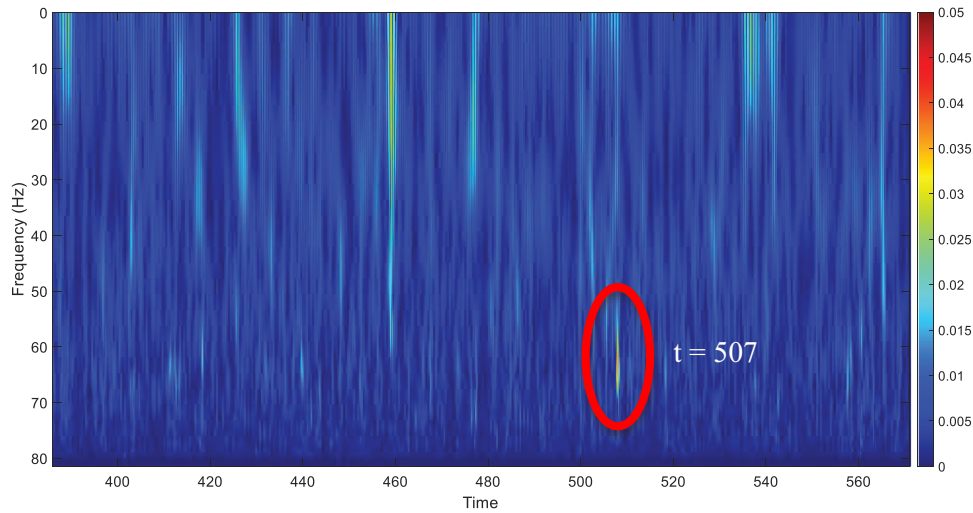

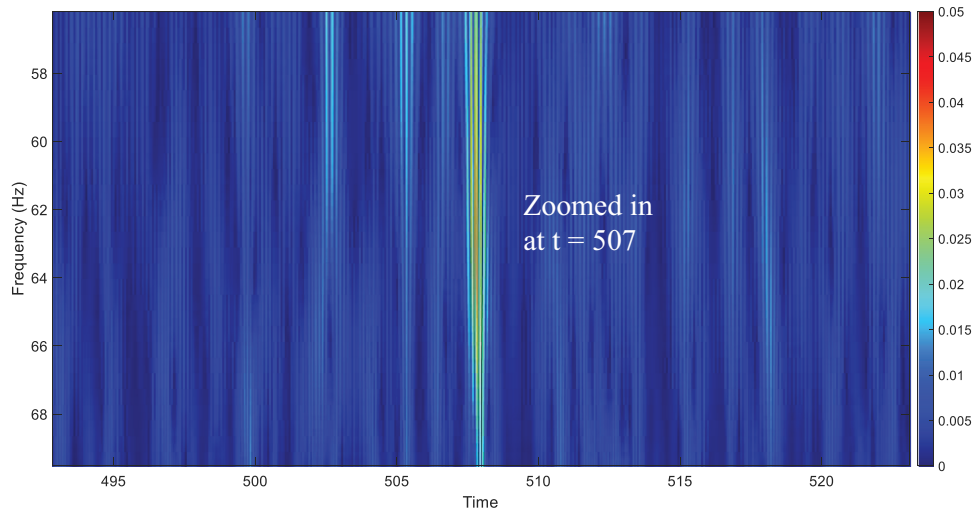

Fig. 17. Wavelet spectrogram for the temperature of case 1 at Re = 5000, Pr = 12, and Ri = 400 on the cold wall side. The energy burst at $t$ = 507 is zoomed in for readability.

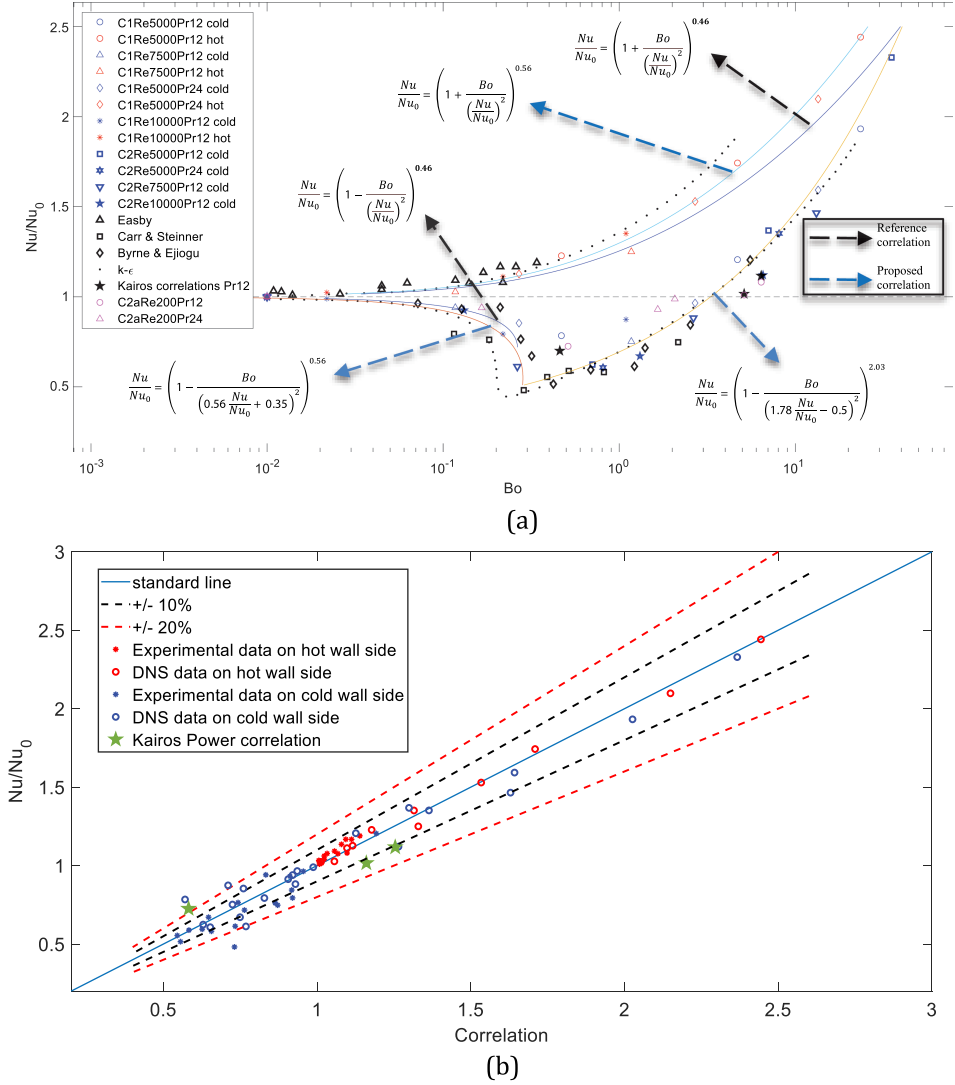

Fig. 18. (a) Nusselt number dataset of simulation cases in Table II as a function of buoyancy number (Bo). Red data points indicate the Nu numbers on the hot wall side, blue data points indicate the Nu numbers on the cold wall side, and black data points indicate the Nu number of reference experiment/KP correlation. (b) Nusselt number line comparison between correlations and experimental, and DNS data.

## VII. CONCLUSION

In this work, we discussed and presented 40 DNS simulations of heated parallel plates representing the downcomer of advanced reactors. The advanced capabilities of the spectral element codes NekRS with DNS are demonstrated. We investigated a wide range of conditions from forced to natural convection corresponding to a set of Re numbers from 5000 to 10 000 and Gr numbers from 0 to 1010. High Pr fluid (FLiBe) has been chosen as the working fluid. Validation of appropriate run-time settings has been acquired by performing a DNS of a low Pr fluid case and then compared with the Kasagi DNS data, and good agreement has been achieved. The run-time settings have then been applied to all simulation cases. Time-averaged turbulence statistics of velocity and temperature for each simulation case have been calculated. The simulation results show that there is a dramatic transition in average velocity and temperature profiles, as well as the rms and TKE budgets, as the Ri number increases. The behavior of turbulence profiles for cases with symmetric cooling BCs is similar to cases with asymmetric cooling on the cold wall side. The TKE budget terms show amplitudes at the hot wall side that are about four times larger than the cold wall side. Furthermore, we observe a negative production region near the cold walls, where the flow is accelerated by the buoyancy force. On the hot wall side, the buoyancy force is directed against the flow, which leads to higher levels of energy production. A strong impact in THF budgets as the Ri number increases is also observed, as the buoyancy force acts on both the hot and cold wall sides.

Changes in the Pr number dramatically affect the flow behavior in this regime. In particular, the amplitude of turbulence statistics decreases as the Pr number grows between 12 and 24, and the bigger Ri is, the stronger this effect will become. In some conditions, the Pr effect can reduce the mixed convection behavior, making mixed convection cases behave nearly like force convection. At the same Pr number but different Re and Ri numbers, mixed convection cases depend largely on the Ri number, whereas the Re number alone has a small impact on the flow behaviors. Finally, we observed a peculiar budget peaking in the axial direction, which has been investigated by implementing POD.

We also collected time series and Nusselt number data. The streamwise velocity time signal at points near the hot wall at a high Ri number shows evidence of low-frequency content associated with high-energy structures. The PSD results show that the highest PSD value is observed in the low-frequency region. Moreover, we observe an energy shift from low to high frequency as buoyancy becomes more dominant: The shift at high frequency occurs at around 20 and 70 Hz. We also observe a temperature PSD peak for case 1.

The wavelet analysis results show that the energy content of wavelet structures increases as the flow transits from forced to natural convection, and the low-frequency behavior between forced and mixed convection cases is revealed in the wavelet spectrogram. We observe a localized wavelet energy peak in the spectrogram that coincides with the frequency range of the PSD peak. Furthermore, the cases that exhibited PSD peaks also showed a significant budget peak in the development region and a strong flow redistribution. This points to the presence of structures that originate in the development region and intermittently propagate downstream. This has never been observed before, to our knowledge.

The assembled Nusselt number dataset for the fully developed region shows a good overall trend agreement with experimental data and available correlations. The behavior of the Nu number in the heat transfer deterioration region is consistent with our observation and the overall decrease in TKE production as the Richardson number increases. A generalized correlation for Nu data in fully developed regions has also been proposed to describe this trend, and the correlation fits most of the experimental, DNS, and KP data within 10% of the confidence interval. In the heat transfer deficient region, the performance of the proposed correlation is less good, and a better correlation is the subject of ongoing work.

**Acronyms**

| | |
|---:|:---|
| BC: | boundary condition |
| Bo: | Buoyancy number |
| CFL: | Courant–Friedrichs–Lewy condition |
| CHT: | conjugate heat transfer |
| DFM: | differential flux model |
| DNS: | direct numerical simulation |
| DOF: | degrees of freedom |
| EB: | elliptic blending |
| Fr: | Froude number |
| GLL: | Gauss-Lobatto-Legendre |
| Gr: | modified Grashof number |
| KP: | Kairos Power |
| KP-FHR: | Kairos Power Fluoride salt-cooled High-temperature Reactor |
| LES: | large eddy simulation |
| NS: | Navier-Stokes |
| OCCA: | Open Concurrent Compute Abstraction |
| OKL: | OCCA kernel language |
| Pe: | Peclet number |
| POD: | proper orthogonal decomposition |
| Pr: | Prandtl number |
| PSD: | power spectral density |
| RANS: | Reynolds-Averaged Navier-Stokes |
| Re: | Reynolds number |
| rms: | root-mean-square |
| SEM: | spectral element method |
| THF: | turbulence heat flux |
| TKE: | turbulence kinetic energy |

## Nomenclature

$c_{p0}$ = inlet FLiBe heat capacity

$D_h$ = hydraulic diameter ($D_h = 2L_y = 4\delta$)

$f$ = dimensional heat flux

$f^*$ = nondimensional heat flux

$k_0$ = inlet FLiBe conductivity

$P_N$ = polynomial order

$P^*$ = nondimensional pressure

$t$ = time

$T$ = temperature of working fluid (FLiBe)

$T_0$ = inlet FLiBe temperature

$T^*$ = nondimensional temperature

$t^*$ = nondimensional time

$U$ = inlet FLiBe velocity

$\vec{u^*}$ = nondimensional vector velocity

*Greek*

$\Delta T$ = temperature difference between inlet and outlet of the bed

$\mu_o$ = inlet FLiBe dynamic viscosity

$\rho_0$ = inlet FLiBe density

$\rho^*$ = nondimensional density

## Disclosure Statement

## Funding

This work was funded by a U.S. Department of Energy Integrated Research Project entitled "Center of Excellence for Thermal-Fluids Applications in Nuclear Energy: Establishing the Knowledgebase for Thermal-Hydraulic Multiscale Simulation to Accelerate the Deployment of Advanced Reactors," IRP-NEAMS -1.1: "Thermal-Fluids Applications in Nuclear Energy." The computational resources are provided by the Oak Ridge Leadership Computing Facility at Oak Ridge National Laboratory.